\documentclass[
	preprint,
	floats,floatfix,superscriptaddress,
	preprintnumbers,showpac,nofootinbib,
]{revtex4-1}
\usepackage{
	amsfonts,amsmath,epsfig,amssymb,latexsym,
	eucal,color,graphicx,array,subfigure,bm,mathrsfs,
	ulem,epsfig,physics,
	mathtools,
}

\newcommand{\pd}{\partial}

\DeclareMathOperator{\Var}{Var}
\DeclareMathOperator{\Cov}{Cov}

\begin{document}

\title{Determining parameters of Kerr black holes \\ at finite distance by shadow observation}
\author{Kenta Hioki}
\email{kenta.hioki@gmail.com}
\address{Sumitomo Mitsui Financial Group, Inc., 1-2, Marunouchi 1-chome, Chiyoda-ku, Tokyo 100-0005, Japan\footnote[1]{The statements expressed in this paper are those of the authors and do not represent the views of Sumitomo Mitsui Financial Group, Inc. or its staff.}}
\author{Umpei Miyamoto}
\email{umpei@akita-pu.ac.jp}
\address{Research and Education Center for Comprehensive Science, Akita Prefectural University, Akita 015-0055, Japan}
\begin{abstract}
We propose a new method to determine the physical parameters of Kerr black holes (namely, specific angular momentum $a$, inclination angle $i$, and distance $D$ from an observer) only from the shadow's information such as the size and shape. Key points in our method are (i) to treat the distance as a free parameter, (ii) to expand the shadow's outline as a Fourier series, and (iii) to construct principal components from the Fourier coefficients. These points enable us to obtain a one-to-one mapping between three principal components, being observables characterizing the size and deviation of shadow's shape from a circular disk, and the values of three parameters ($D/M, a/M, i$), where $M$ is the mass of black hole. Our method is applicable to various type of black holes and even to ones with accretion disks.
\end{abstract}

\maketitle

\tableofcontents

\section{Introduction}
\label{sec:intro}
A black hole is an intriguing celestial object and serves as the ultimate testing ground for the physics of strong gravitational fields. Observing the shadow of a black hole has been a goal in physics for a long time. The shadow forms when the path of light emitted from the matter surrounding the black hole is curved by the gravitational pull of the black hole itself.

Recently, Earth-sized very long baseline interferometers successfully captured images of two black hole candidates. The first image depicted M87 \cite{EventHorizonTelescope:2019dse, EventHorizonTelescope:2019uob, EventHorizonTelescope:2019jan, EventHorizonTelescope:2019ths, EventHorizonTelescope:2019ggy, EventHorizonTelescope:2021bee, EventHorizonTelescope:2021srq}, while the second featured ${\rm Sgr~A}^\ast$ \cite{EventHorizonTelescope:2022xnr, EventHorizonTelescope:2022vjs, EventHorizonTelescope:2022wok, EventHorizonTelescope:2022exc, EventHorizonTelescope:2022urf, EventHorizonTelescope:2022xqj}. Both images showcased a prominent ring, indicating the successful establishment of a technique for capturing black hole images. However, the shadow, representing the darker portion of the image, has not yet been identified. Further enhancements in observation equipment are needed to observe these shadows~\cite{EventHorizonTelescope:2019ths}.

Some researchers argue that the exploration of black hole imaging traces back to the derivation of the shadow contour, now referred to as the apparent shape. The apparent shape of a Schwarzschild black hole was initially derived in Ref.~\cite{darwin1959gravity}, and that of a Kerr black hole was established in Ref.~\cite{Bardeen:1973xx}. These pioneering works form the foundation of shadow theory. Although these calculations focused on determining the apparent shapes of simple ``bare'' black holes, excluding considerations of accretion disks around them, they unveiled the crucial role of the photon sphere. This element remains central even in the shadow of a black hole with an accretion disk.

Numerous researchers have delved into the study of black hole shadows and techniques for extracting physical information, such as the angular momentum of the black hole, through shadow observations~\cite{Hioki:2008zw, Bambi:2010hf, Amarilla:2010zq, Amarilla:2013sj, Wei:2013kza, Papnoi:2014aaa, Wei:2015dua, Singh:2017vfr, Stuchlik:2019uvf}. However, our focus in this paper is to underscore that there is still room for enhancing the determinability of the physical parameters of black holes through shadow analysis. Specifically, further investigation is needed to ascertain whether extracting information, such as the angular momentum of the black hole, solely from shadow observations is possible.

One approach to explore this possibility is to investigate whether a map from a parameter space to an image library is an injection~\cite{Hioki:2009na, Hioki:2022mdg}. It has been demonstrated that the map from the parameter space to the apparent-shape library for a bare Kerr black hole is indeed a bijection~\cite{Hioki:2009na}. This result implies that the specific angular momentum (per mass) and inclination angle of the Kerr black hole can be determined by observing its apparent shape. Here, the apparent-shape library is defined as the set of all possible apparent shapes that can be generated in the given gravitational theory and model.

Confirming (1) whether an object is a black hole, (2) identifying the specific black hole solution, and (3) determining its physical quantities solely through the observation of the shadow image of a black hole candidate is a complex question that demands extensive research. It is imperative to generate a diverse array of models for relativistic objects, thereby constructing an apparent-shape library with varying parameters. Additionally, an investigation into whether the map is injective (or invertible) is crucial. If the map is not injective, then it signifies the existence of an image corresponding to different models and parameter settings, thereby making it impossible to determine the model based on shadow observation alone. These research efforts are currently underway, and further studies are warranted.

The quantities that characterize the apparent shape are referred to as observables. The observables proposed to determine the parameters of the Kerr black hole have been applied to other black hole solutions. Some attempts have also been made to enhance these observables~\cite{Abdujabbarov:2015xqa}. The idea was to expand the apparent shape in terms of Legendre polynomials. When dealing with systems with a larger number of parameters, intuitively found observables may not be sufficient.

It is imperative to establish observables characterizing the apparent shape of relativistic objects, not only for black holes. If these observables can distinguish all elements in the apparent-shape library, then determining the system's parameters becomes achievable through shadow observations.

This paper introduces a systematic method for characterizing apparent shapes. Initially, the apparent shape undergoes Fourier expansion, followed by transforming the Fourier coefficients into principal components via principal component analysis. These principal components serve as observables. Principal component analysis is a standard method for reducing multivariable data to lower dimensions, and the resulting data are known as principal components.

The systematic method is applied to Kerr black holes, treating the black hole's distance as a finite free parameter. The system involves parameters such as distance $D$, specific angular momentum $a$, inclination angle $i$, and mass $M$. Consequently, we show that the map from parameter space to the apparent-shape library is injective. The system's parameters ($D/M$, $a/M$, $i$) can be exclusively determined through shadow observations.

Furthermore, we proof the uniqueness concerning apparent shapes within the apparent-shape library. Uniqueness is defined as the absence of two congruent apparent shapes for two distinct parameter values.

The organization of this paper is as follows. In Sec.~\ref{sec:setup}, we prepare for the analysis with writing down the null geodesic equations around the Kerr black hole, describing how to draw a two-dimensional image from the geodesics, and we define a map from a space of physical parameters to a space of two-dimensional images. In Sec.~\ref{sec:shadow}, we present several examples of apparent shapes and then prove analytically the uniqueness of apparent shapes. In Sec.~\ref{sec:pc}, we propose a new method to systematically construct observables, which is applied to Kerr black holes at finite distances, and this shows that the parameters of the system can be determined from shadow observations. We summarize our analysis and give future prospects in the final section. In the Appendix, we present the details of the principal component analysis and specify the vectors and matrices needed to obtain the principal components as observables. We use the geometrized units, where $c=G=1$.

\section{Setup}
\label{sec:setup}

\subsection{Null geodesics, light source, and observer}
The spacetime of a rotating black hole is believed to be well described by the Kerr metric~\cite{Kerr}, which is given by
\begin{eqnarray}
	g_{\mu \nu}{\rm d}x^\mu {\rm d}x^\nu 
	&=&
	-\left(
		1-\frac{2Mr}{\varSigma}
	\right) {\rm d}t^2
	+
	\frac{\varSigma}{\varDelta} {\rm d}r^2
	+
	\varSigma {\rm d}\theta ^2
	-\frac{ 4Mra\sin^2\theta }{ \varSigma } {\rm d}t {\rm d}\phi
	+
	\frac{A \sin ^2 \theta}{\varSigma} {\rm d}\phi ^2
	,
	\label{eq:metric}
\end{eqnarray}
where $x^\mu = \left( t, r, \theta , \phi \right)$ $\left( \mu , \nu = 0, 1, 2, 3 \right)$,
\begin{eqnarray}
    \varSigma (r,\theta)
	:=
	r^2 + a^2 \cos^2 \theta,
\;\;
    \varDelta (r)
	:=
	r^2 - 2Mr + a^2,
\;\;
    A (r,\theta)
	:=
	\left( r^2 + a^2 \right) ^2 - a^2 \varDelta \sin^2\theta.
\end{eqnarray}
The parameters $M$ and $a$ are the mass and specific angular momentum of spacetime, respectively. If $\left| a \right| \leq M$, then an event horizon exists in spacetime, and the metric describes a black hole. The radii of the outer and inner horizons are given by $r_{+} = M + \sqrt{M^2 -a^2}$ and $r_{-} = M - \sqrt{M^2 -a^2}$, respectively.

To find the trajectory $x^\mu \left( \lambda \right)$ of a massless test particle as a light ray, we shall consider the geodesic equation. Since the Kerr spacetime admits four constants of motion in involution, the geodesic equation is completely integrable.

Starting from Lagrangian,
\begin{eqnarray}
	\mathcal{L}
	=
	\frac{1}{2}g_{\mu \nu}\dot{x}^\mu \dot{x}^\nu
		, \;\;\; \dot{x}^\mu := \frac{dx^\mu}{d\lambda} ,
	\label{eq:lag}
\end{eqnarray}
we deduce the energy $E$ and the axial component of the angular momentum $L$ of the test particle:
\begin{eqnarray}
	E \coloneqq - \frac{\partial \mathcal{L}}{\partial \dot{t}} = \left( 1-\frac{2Mr}{\varSigma} \right) \dot{t} +\frac{ 2Mra\sin^2\theta }{ \varSigma } \dot{\phi}
\end{eqnarray}
and
\begin{eqnarray}
	L \coloneqq \frac{\partial \mathcal{L}}{\partial \dot{\phi}}  = -\frac{ 2Mra\sin^2\theta }{ \varSigma } \dot{t} + \frac{A \sin ^2 \theta}{\varSigma} \dot{\phi} \, .
	\label{eq:lag}
\end{eqnarray}
Here, quantities $\mathcal{L}$, $E$, and $L$ are the constants of motion, and so is $Q$, being called the Carter constant~\cite{Chandrasekhar:1985kt}.

Introducing two conserved quantities $\ell \coloneqq L/E$ and $q \coloneqq Q/E^2$ for null geodesics $(\mathcal{L} = 0)$, we have the following set of first-order differential equations:
\begin{eqnarray}
	&&
 	\varSigma \frac{{\rm d} t}{{\rm d}\tilde{\lambda}}
	=
	\frac{A-2Mra \ell }{\varDelta} \, ,
	\label{eq:velocity2}	
	\\
	&&
	\varSigma \frac{{\rm d} r}{{\rm d}\tilde{\lambda}}
	=
	\pm \sqrt{R} \, ,
	\label{eq:velocity0}
	\\
	&&
	\varSigma \frac{{\rm d} \theta}{{\rm d}\tilde{\lambda}}
	=
	\pm \sqrt{\varTheta} \, ,
	\label{eq:velocity1}
	\\
	&&
	\varSigma \frac{{\rm d} \phi}{{\rm d}\tilde{\lambda}}
	=
	\frac{ 2Mar+\ell \csc ^2\theta (\varSigma - 2Mr) }{\varDelta} \, ,
	\label{eq:velocity3}
\end{eqnarray}
where the affine parameter $\tilde{\lambda}$ is defined as $\tilde{\lambda} \coloneqq \lambda E^2$ and
\begin{gather}
	k  := q +\left( a-\ell \right) ^2 \, ,
	\;\;
	R(r)
	:=
        \left( r^2+a^2-a\ell \right) ^2-k \varDelta \, ,
	\;\;
	\varTheta(\theta)
	:=
	k -(a \sin \theta -\ell \csc \theta )^2.
\label{eq:potential}
\end{gather}

We would like to describe the setting of the observer. Candidate observers in Kerr spacetime are zero-angular-momentum observers~\cite{Bardeen:1973xx} and Carter's observers~\cite{Grenzebach:2014fha}. These two observers have different motions in the azimuthal direction~\cite{Chang:2020lmg}. The aim of this paper is to show the existence of an observer who can determine the black hole parameters. It is sufficient to choose one from either observer. In this paper, the latter is adopted as the observer.

We choose a tetrad basis:
\begin{gather}
	e_{(t)} = \frac{\left( r^2 + a^2 \right) \pd_t + a \pd_\phi }{\sqrt{\varSigma \varDelta}} \, ,
\;\;\;
	e_{(r)} = - \sqrt{\frac{\varDelta}{\varSigma}} \pd_r \, ,  \\
	e_{(\theta)} = \frac{1}{\sqrt{\varSigma}} \pd_\theta \, ,
\;\;\;
	e_{(\phi)} = -\frac{ a \sin \theta \pd_t + \csc \theta \pd_\phi }{\sqrt{\varSigma}}  \, .
\label{eq:tetrad}
\end{gather}
Note that $e_{(t)} | _{(r, \theta ) = (D, i)}$ is also the four-velocity of the observer [see Fig.~\ref{fig010}(a)]. For the observer, $e_{(r)} | _{(r, \theta ) = (D, i)}$ gives the spatial direction towards the black hole.

In many studies, observers have been assumed to be at spatial infinity. Since the actual distance between the black hole and the observer is finite, we assume that the observer is at finite distance $D$ from the black hole. Although the observer can exist anywhere in the domain of outer communication, we restrict the distance to the following range: $D \in \left[ 10M , 50M \right]$, which is sufficient to show that the parameters can be determined. The method of determining the parameters described in this paper is not dependent on this distance limitation.

In general, the image of a black hole varies with the surrounding conditions such as the presence or absence of an accretion disk and the location of the light source. We are interested in considering the shadow of an astronomical black hole.

The following assumptions are appropriate when considering the shadow of an astronomical black hole such as M87~\cite{EventHorizonTelescope:2021dqv}. We assume there is no gas accretion flow or accretion disk around the black hole, i.e., the black hole is bare. Then, we assume that the light sources are distributed on $r=r_e = {\rm const.}$, where $r_e>D$~\cite{Grenzebach:2014fha}.

\subsection{How to define a two-dimensional image}
\label{sec:shadow}
\begin{figure}[tb]
		\begin{tabular}{ cc }
			\includegraphics[height=5.5cm]{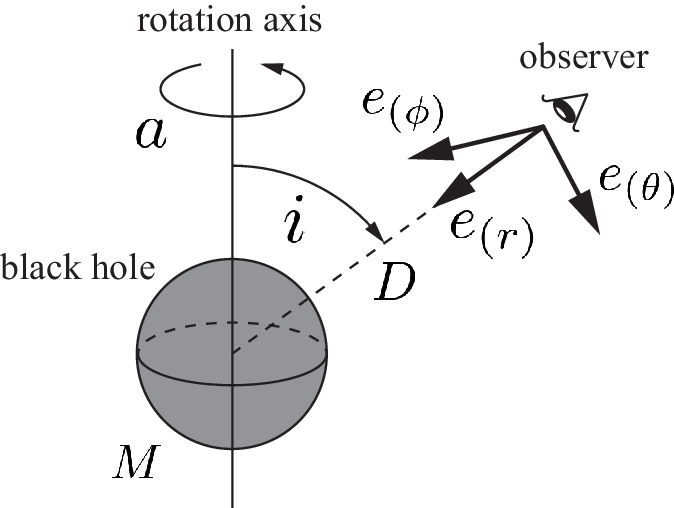} &
			\includegraphics[height=5.5cm]{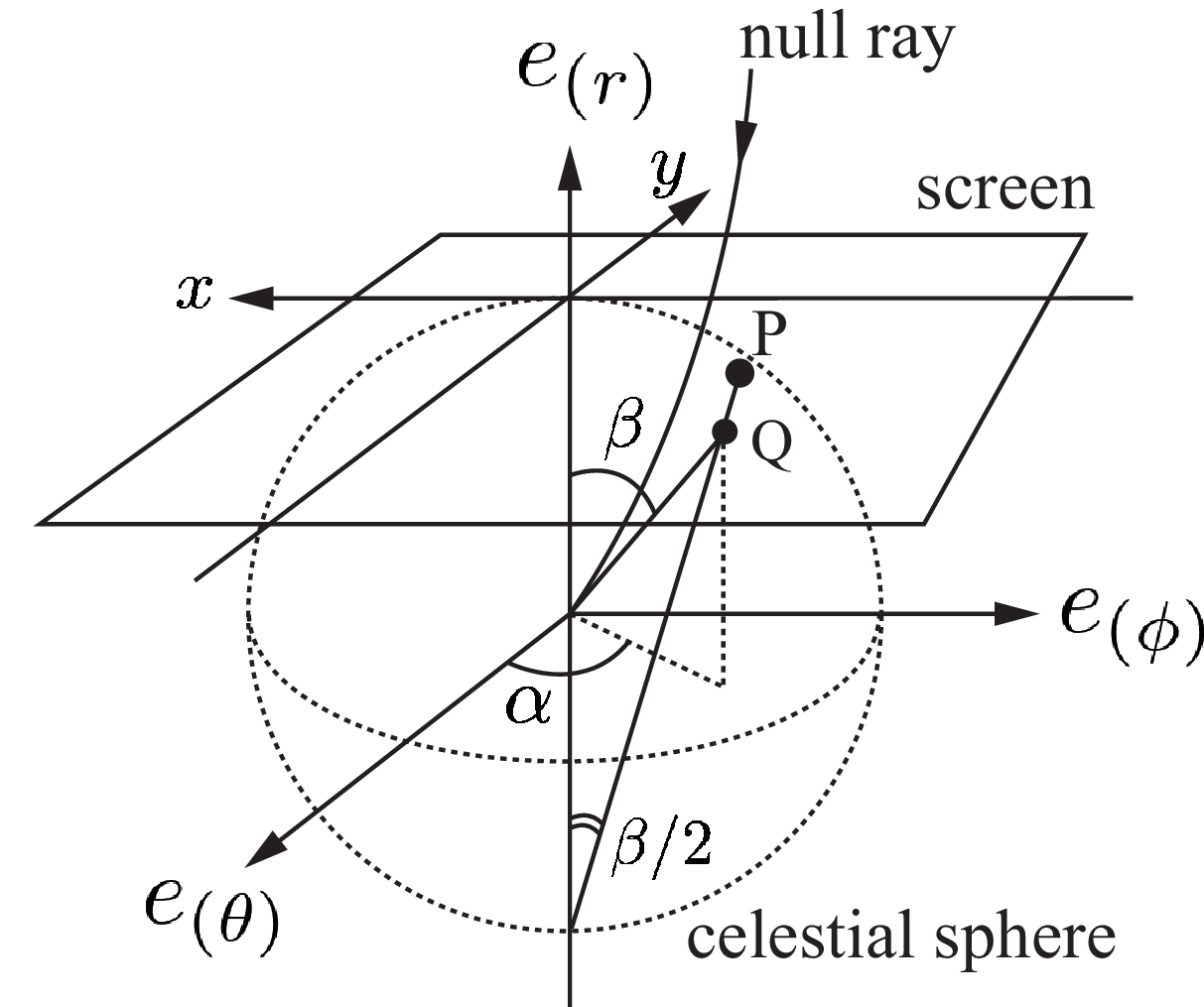}
   \\
			(a) & (b)
		\end{tabular}
	\caption{\footnotesize{
(a) A schematic picture showing a unit sphere in which the center coincides with that of the Kerr black hole. The observer is located at $(r, \theta) = (D, i)$. (b) A schematic picture showing a unit celestial sphere for the observer. The direction of $e_{(r)}$ is a direction to the Kerr black hole from the observer. $(\alpha , \beta)$ is a celestial coordinate system, which is used to specify the incident angle of the null ray into the observer. Q is the point where the tangent of null ray at the center of the celestial sphere crosses the unit sphere. P is the two-dimensional image made by the null ray.}}
	\label{fig010}
\end{figure}
We denote the tangent vector of a null geodesic $x^\mu(\tilde{\lambda})$ by
\begin{eqnarray}
	\frac{{\rm d}}{{\rm d} \tilde{\lambda}} = \frac{{\rm d} t}{{\rm d}\tilde{\lambda}} \partial _t + \frac{{\rm d} r}{{\rm d}\tilde{\lambda}} \partial _r + \frac{{\rm d} \theta}{{\rm d}\tilde{\lambda}} \partial _\theta + \frac{{\rm d} \phi}{{\rm d}\tilde{\lambda}} \partial _\phi \, .
\label{eq:tangentv1}
\end{eqnarray}
From Fig.~\ref{fig010}(b), this tangent vector at the observer can be written using two incident angles as
\begin{eqnarray}
	\frac{{\rm d}}{{\rm d} \tilde{\lambda}} = \epsilon \left( - e_{(t)} + \cos \beta e_{(r)} + \sin \beta \cos \alpha e_{(\theta)} + \sin \beta \sin \alpha e_{(\phi)} \right) ,
\label{eq:tangentv2}
\end{eqnarray}
where $\epsilon$ is a scalar factor. We call these incident angles $\alpha$ and $\beta$ the observer's celestial coordinates. Since the basis vectors are orthonormal, then
\begin{eqnarray}
	\epsilon = \left. \left( \frac{{\rm d}}{{\rm d}\tilde{\lambda}} \right) ^{\mu} e_{(t)\mu} \right| _{(r, \theta) = (D, i)} = - \left. \frac{ r^2 + a^2 - a\ell}{\sqrt{\varSigma \varDelta}} \right| _{(r, \theta) = (D, i)} \, .
\label{def_I}
\end{eqnarray}

Using Eqs.~(\ref{eq:tangentv1}) and (\ref{eq:tangentv2}), we can derive the following relations between the celestial coordinates and the conserved quantities:
\begin{eqnarray}
\begin{split}
	&&
	\sin \alpha
	=
	\left. \frac{\sin \theta}{\sqrt{\varDelta}\sin \beta} \left( \frac{\varSigma \Delta}{r^2 + a^2  - a \ell} \frac{{\rm d} \phi}{{\rm d}\tilde{\lambda}} -a \right) \right| _{(r, \theta) = (D, i)} \, ,
	\\
	&&
	\cos \beta
	=
	\left. \frac{\varSigma}{r^2 + a^2 - a \ell} \frac{{\rm d} r}{{\rm d}\tilde{\lambda}} \right| _{(r, \theta) = (D, i)} \, .
\end{split}
	\label{eq:alphabeta}
\end{eqnarray}

In Fig.~\ref{fig010}(b), we introduce screen coordinates $(x, y)$~\cite{Grenzebach:2014fha} to draw the image of a black hole on a two-dimensional plane. We map celestial coordinates $(\alpha , \beta )$ to screen coordinates $(x, y)$ as follows:
\begin{eqnarray}
	x
	=
	-2 \sin \alpha \tan \frac{\beta}{2}  \, ,
\;\;\;
	y
	=
	-2 \cos \alpha \tan \frac{\beta}{2} \, .
	\label{eq:xy}
\end{eqnarray}
A circle with radius 2 in the $x$-$y$ plane corresponds to the celestial equator.

We now consider the geodesics that form the apparent shape of the black hole. Whether light emitted from a source has a turning point and can reach an observer depends on the value of the conserved quantities. The critical conserved quantities are those of geodesics that wind around an unstable spherical photon orbit an infinite number of times.
 
A geodesic with $r={\rm const.}$ is called a {\it spherical photon orbit}. The radius of such an orbit $r_s$ is given by
\begin{eqnarray}
	R(r_s)
	 = 
	0  \, ,
\;\;\;
	R'(r_s)
	 = 
	0 \, 
	\label{eq:spo}
\end{eqnarray}
with the additional condition that there exists some interval of $\theta$ within $[0, \pi]$ in which 
\begin{eqnarray}
	\varTheta (\theta)
	& \geq &
	0  \, 
	\label{eq:contheta}
\end{eqnarray}
holds. The spherical photon orbits form a one-parameter family. We denote $\ell$ and $q$ satisfying Eq.~\eqref{eq:spo} by
\begin{eqnarray}
	\ell _s
	 = 
	\frac{r_s^2+a^2}{a} - \frac{2r_s\varDelta (r_s)}{a(r_s-M)} \, ,
\;\;\;
	q_s
	 = 
	- \frac{r_s^3[r_s(r_s-3M)^2-4a^2M]}{a^2(r_s-M)^2} \, ,
	\label{eq:spoxieta}
\end{eqnarray}
respectively.
Note that $\ell _s$ and $q_s$ are two conserved quantities of the spherical photon orbits. Inserting Eq.~(\ref{eq:spoxieta}) into Eq.~(\ref{eq:potential}), we find
 \begin{eqnarray}
	k=k_s:= \frac{4r_s^2\varDelta(r_s) }{(r_s - M)^2}  \, .
	\label{eq:cartercnst}
\end{eqnarray}
Since $k$ is non-negative in general~\cite{Chandrasekhar:1985kt}, the radius of the spherical photon orbit $r_s$ must be in the range of $r_s \in ( -\infty , r_{-}] \cup [r_{+}, \infty )$. This condition is a necessary condition for the existence of a spherical photon orbit. As can be seen from Eq.~(\ref{eq:contheta}), the radius $r_s$ must also satisfy $\varTheta (i) | _{(\ell , q) = (\ell _s , q_s)} \geq 0$.

The condition for the spherical photon orbit to be unstable is
\begin{eqnarray}
    R''(r_s)
    > 0 \, ,
	\label{eq:sta}
\end{eqnarray}
which is equivalent to $r_s <0$ or $M-[M(M^2-a^2)]^{1/3}<r_s$~\cite{Hioki:2009na}. Therefore, assuming there exists a spherical orbit at the radius $r_{s}$, we find that the condition for the orbit to be unstable is
\begin{eqnarray}
	r_{s} \in ( -\infty , 0 ) \cup ( r_{+} , \infty  )  \, .
	\label{eq:rsta}
\end{eqnarray}
All spherical photon orbits outside the event horizon are unstable. We define $I$ as the set of values representing the radii of existing unstable spherical photon orbits.

We clarify the relation between the apparent shape and the unstable spherical photon orbits. Let us trace the geodesics that form the apparent shape backward from the end point, the observer. The geodesics ending at the observer can be divided into those starting from the light source and those not starting from the light source. Furthermore, geodesics that do not start from a light source can be divided into those that start from a black hole and those that are infinitely wound around an unstable spherical photon sphere.

Let us draw a curve on the screen coordinates corresponding to the geodesics that are infinitely wound around the unstable sphere orbit. The functions as the correspondence from a set of geodesics to a set of points on the screen coordinates are $x(\ell _s, q_s)$ and $y(\ell _s, q_s)$ which are obtained by substituting Eqs.~(\ref{eq:alphabeta}), (\ref{eq:velocity0}), and (\ref{eq:velocity3}) into Eq.~(\ref{eq:xy}). The functions can be reduced to $x(r_s)$ and $y(r_s)$ by using Eq.~(\ref{eq:spoxieta}). We can draw a closed curve $c$ on the screen coordinates for a fixed value of $(M, D, a, i)$ by changing $r_s$ within the whole range of $I$.

Since the spherical photon orbits are unstable, the boundary between dim and bright regions on the screen coordinates is the closed curve $c$~\cite{Hioki:2022mdg, Hioki:2009na}. The geodesic that starts from the black hole and reaches the observer straight ($\theta = {\rm const.}$) with no turning point, principal null geodesic, corresponds to a point inside the closed curve $c$ on the screen coordinates~\cite{Hioki:2009na}. This point is part of the shadow because it is not a geodesic from a light source. Then, geodesics corresponding to the inside and outside of the closed curve $c$ can be said to have started from the black hole and the light source, respectively. The closed curve $c$ itself is, of course, a part of the shadow. Finally, the closed curve $c$ in the screen coordinates is the contour of the apparent shape of the black hole.

\subsection{Map from parameter space to apparent-shape library}
\label{sec:map}
To examine whether one can determine the parameters of the system by observation, it is convenient to define a parameter space $P$, an apparent-shape library $S$, and a map $f_1 : P \rightarrow S$. The parameter space in our problem is defined as follows:
\begin{eqnarray}
	P &\coloneqq& \left\{ \left( M, D, a, i \right) \biggm\vert M > 0, 10 \leq \frac{D}{M} \leq 50 , 0 < \frac{a}{M} \leq 1, 0 \leq i \leq \frac{\pi}{2} \right\} \cup P_0 \subset \mathbb{R}^4  \, ,
	\label{eq:pspace}
\end{eqnarray}
where
\begin{eqnarray}
	P_0 \coloneqq \left\{ \left( M, D, 0, 0 \right) \biggm\vert M > 0, 10 \leq \frac{D}{M} \leq 50 \right\} \, .
\end{eqnarray}
When the specific angular momentum $a$ is zero, the black hole is spherically symmetric and has no axis of rotation, so the inclination angle $i$ need not be considered. The subspace $P_0$ corresponds to the case where the black hole is a Schwarzschild black hole.

We also define an equivalence relation $\sim$ in the parameter space $P$:
\begin{eqnarray}
	\left( M, D, a, i \right) \sim \left( M', D', a', i' \right)
	\stackrel{\rm def.}{\iff}
	\begin{cases}
		D/M = D'/M' \\
		a/M = a'/M' \\
		i = i' 
	\end{cases} .
\end{eqnarray}
The apparent-shape library $S$ is defined as the set of all apparent shapes generated by the system. The map $f_1$ from the parameter space $P$ to the apparent-shape library $S$ is defined as
\begin{eqnarray}
	f_1 : P \ni \left( M, D, a, i \right) \mapsto \{ \left. \left( x(r_s) , y(r_s) \right) \right| r_s \in I \} \in S ,
\end{eqnarray}
where $x(r_s)$ and $y(r_s)$ are the functions described in the previous section. 

From the definitions, the map $f_1$ is surjective, $S = f_1 \left( P \right)$. It is not obvious whether the map is injective or not. If the map is injective, then one element in the apparent-shape library $S$ corresponds to one element of the parameter space $P$. In other words, if we can observe the apparent shape of a black hole, we can determine its parameters, in principle. In Sec.~\ref{sec:pc}, we will go deeper into the parameter determinability.

\section{Analysis of size and shape of shadow}
\label{sec:shadow}

\subsection{Examples of shadow}
\label{subsec:examples}

\begin{figure}[tb]
		\begin{tabular}{ cccc }
			\includegraphics[height=3.8cm]{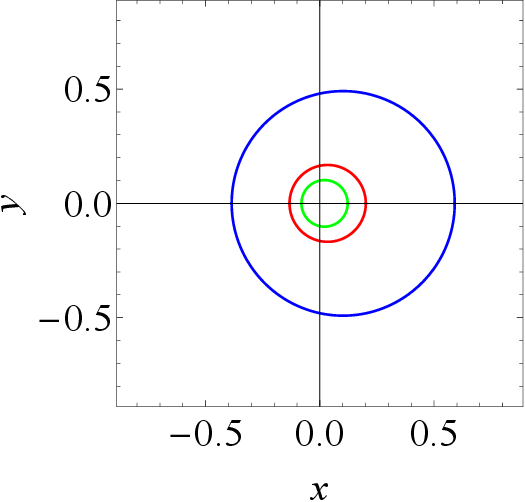} &
			\includegraphics[height=3.8cm]{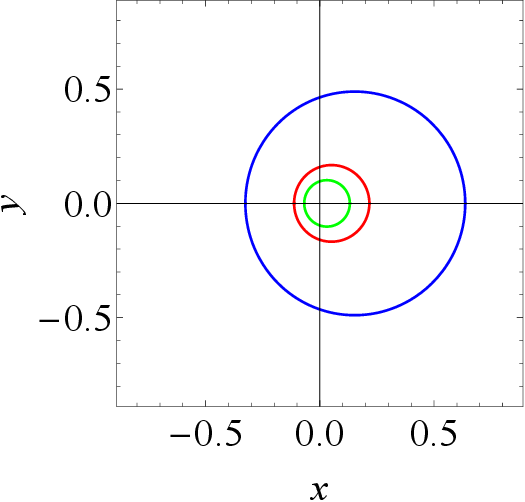} &
			\includegraphics[height=3.8cm]{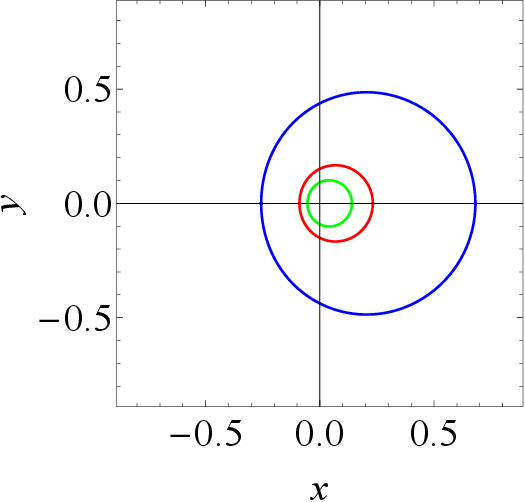} &
			\includegraphics[height=3.8cm]{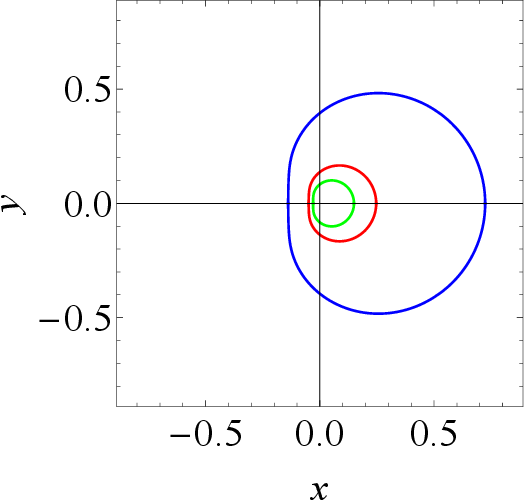} \\
			(a) $a/M=0.4$, $i=60^\circ$ & (b) $a/M=0.6$, $i=60^\circ$ & (c) $a/M=0.8$, $i=60^\circ$ & (d) $a/M=0.999$, $i=60^\circ$ \\
			\includegraphics[height=3.8cm]{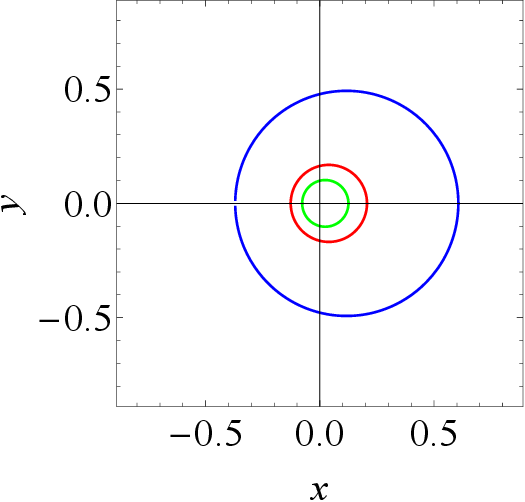} &
			\includegraphics[height=3.8cm]{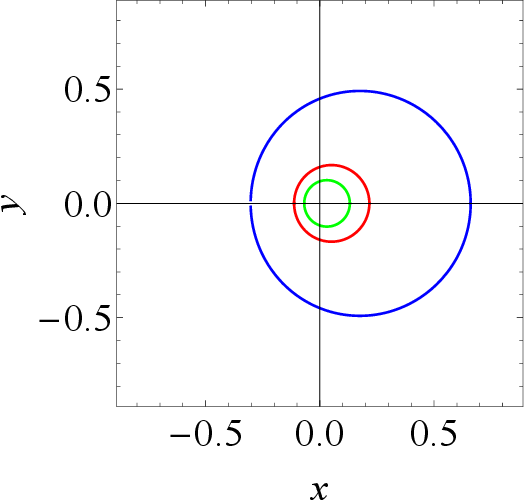} &
			\includegraphics[height=3.8cm]{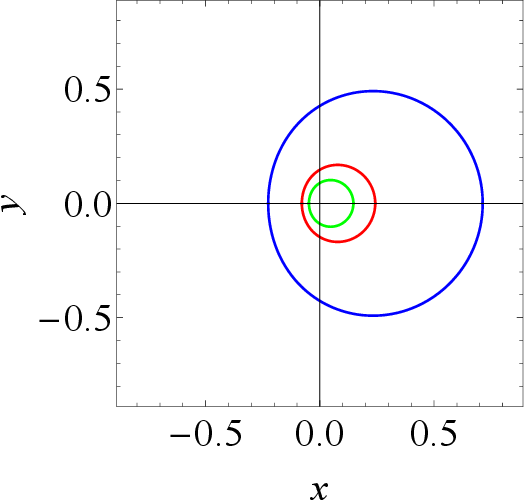} &
			\includegraphics[height=3.8cm]{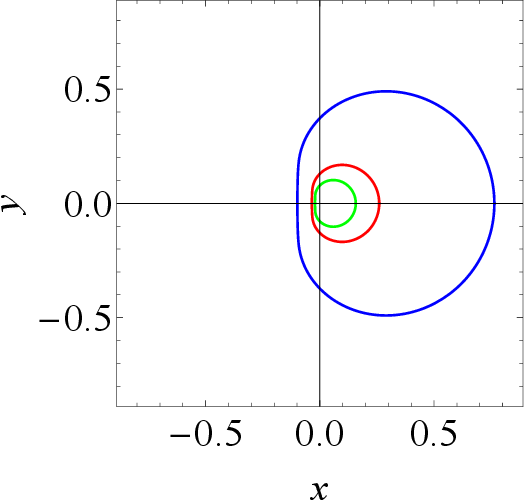} \\
			(e) $a/M=0.4$, $i=90^\circ$ & (f) $a/M=0.6$, $i=90^\circ$ & (g) $a/M=0.8$, $i=90^\circ$ & (h) $a/M=0.999$, $i=90^\circ$
		\end{tabular}
	\caption{\footnotesize{Shadows of the Kerr black holes. The blue, red, and green curves are the apparent shape of the black hole at dimensionless distances $D/M$ of 10, 30, and 50, respectively.}}
	\label{fig020}
\end{figure}
In Fig.~\ref{fig020}, we present the apparent shapes of the Kerr black holes for four values of dimensionless spin parameter $a/M = 0.4$, 0.6, 0.8, and 1, two values of inclination angle $i = 60^\circ $ and $90^\circ $, and three values of dimensionless distance $D/M = 10$, 30, and 50.

The parameters $a/M$ and $i$ mainly change the shape of the shadow, while the parameter $D/M$ mainly changes the size of the shadow. As we will see in Sec.~\ref{sec:pc}, $D/M$ also changes the shape of the shadow. In other words, the apparent shapes of black holes with different $D/M$ are {\it not similar}. (One shape cannot be obtained from the other by either the uniform scaling, translation, rotation, or reflection.)

\subsection{Uniqueness of apparent shape}
\label{sec:prop}
We now introduce the following dimensionless parameters:
\begin{eqnarray}
	D_\ast := \frac{D}{M}, \;\;\;
	a_\ast := \frac{a}{M}, \;\;\;
	r_{\ast} := \frac{r_s}{M}.
	\label{eq:Dar}
\end{eqnarray}
Then, we see that the image on the screen coordinates is drawn by a function of the dimensionless parameters $(r_{\ast}, D_\ast, a_\ast, i)$ and does not depend on $M$ explicitly. In other words, all elements of a particular equivalence class $\left[ \left( M, D, a, i \right) \right]$ in the quotient space $P/\sim$ are mapped to one element in the apparent-shape library by a map $\tilde{f}_1$. Here, the map is defined as $\tilde{f}_1: P/\sim \ni [(M, D, a, i)] \mapsto f_1((M, D, a, i)) \in S$.

Here, let us show that the map $\tilde{f}_1$ is injective, and an apparent shape of a bare Kerr black hole is unique. We refer to uniqueness as the absence of two congruent apparent shapes for two values of parameters $\left( D_\ast , a_\ast , i \right)$. To show this, it is essential to analyze $\sin \alpha$ and $\cos \beta$ in Eq.~\eqref{eq:alphabeta}.

We find that $\sin ^2 \alpha$ is a rational function of $r_\ast = r_s/M$. In the case of the nonextreme Kerr black hole ($0 \leq |a| <M$), the numerator and denominator of $\sin ^2 \alpha$ do not have a common factor. Therefore, $\sin ^2 \alpha$ is irreducible. In the case of the extreme Kerr black hole ($|a|=M$), $(r_\ast -1)^2$ is a common factor between the numerator and denominator in $\sin ^2 \alpha$. Dividing the numerator and denominator of $\sin ^2 \alpha$ by the common factor in the latter case, $\sin ^2 \alpha$ becomes irreducible:
\begin{eqnarray}
	\sin ^2 \alpha
	=
	\begin{cases}
	\frac{\sum _{n=0}^{6} a_n r_\ast^n}{\sum _{n=0}^{4} b_n r_\ast^n} & (0 \leq |a| <M) \\
	\frac{\sum _{n=0}^{4} \bar{a}_n r_\ast^n}{\sum _{n=0}^{2} \bar{b}_n r_\ast^n} & (|a|=M)
	\end{cases} ,
\label{eq:sinalpha-nonext}
\end{eqnarray}
where
\begin{gather*}
	a_0 = 4 a_\ast^4 \cos ^4 i,
\;\;\;
	a_1 = -4 a_\ast^2 \cos ^2 i \left( \cos 2i -3 \right),
\\
	a_2 = a_\ast^2 \left[  -12 + 9a_\ast^2 -6 \left(2 + a_\ast^2 \right) \cos 2i +a_\ast^2 \cos ^2 2i   \right],
\\
	a_3 = 16a_\ast^2 (\cos 2i -2),
\;\;\;
	a_4 = 12(3+a_\ast^2) -4a_\ast^2 \cos 2i,
\;\;\;
	a_5 = -24,
\;\;\;
	a_6 = 4,
\\
	b_0 = 0,
\;\;\;
	b_1 = 0,
\;\;\;
	b_2 = 16 a_\ast^4 \sin ^2 i,
\;\;\;
	b_3 = -32a_\ast^2 \sin ^2 i, 
\;\;\;
	b_4 = 16 a_\ast^2 \sin ^2 i
\end{gather*}
and
\begin{gather*}
	\bar{a}_0 = 1 + 2 \cos 2i + \cos ^2 2i,
\;\;
	\bar{a}_1 = 8+8 \cos 2i,
\;\;
	\bar{a}_2 = 12 -4 \cos 2i,
\;\;
	\bar{a}_3 = -16,
\;\;
	\bar{a}_4 = 4,
\\
	\bar{b}_0 = 0,
\;\;\;
	\bar{b}_1 = 0,
\;\;\;
	\bar{b}_2 = 16.
\end{gather*}

We find that $\cos ^2 \beta$ is also a rational function of $r_\ast$. In the case of the nonextreme Kerr black hole, the numerator and denominator of $\cos ^2 \beta$ do not have a common factor. Hence, $\cos ^2 \beta$ is irreducible. In the case of the extreme Kerr black hole, $(r_\ast -1)^2$ is a common factor between the numerator and denominator of $\cos ^2 \beta$. Dividing the numerator and denominator of $\cos ^2 \beta$ by the common factor in the latter case, $\cos ^2 \beta$ becomes irreducible:
\begin{eqnarray}
	\cos ^2 \beta
	=
	\begin{cases}
	\frac{\sum _{n=0}^{6} e_n r_\ast^n}{\sum _{n=0}^{6} g_n r_\ast^n} & (0 \leq |a| <M) \\
	\frac{\sum _{n=0}^{4} \bar{e}_n  r_\ast^n}{\sum _{n=0}^{4} \bar{g}_n r_\ast^n} & (|a|=M)
	\end{cases}
	,
	\label{eq:cosbeta-nonext}
\end{eqnarray}
where
\begin{gather*}
	e_0 = D_\ast^4,
\;\;\;
	e_1 = -2 \left( 2a_\ast^2 D_\ast^2 +D_\ast^4 \right),
\;\;\;
	e_2 = D_\ast \left(8a_\ast^2 +6D_\ast + D_\ast^3 \right),
\\
	e_3 = -4\left( a_\ast^2 +4D_\ast \right),
\;\;\;
	e_4 = 9 -2 D_\ast (D_\ast-4),
\;\;\;
	e_5 = -6,
\;\;\;
	e_6 = 1,
\\
	g_0 = D_\ast^4,
\;\;\;
	g_1 = -2 \left(2a_\ast^2 D_\ast^2 + D_\ast^4 \right),
\;\;\;
	g_2 = 6D_\ast^2 + \left( 2a_\ast^2 + D_\ast^2 \right)^2,
\\
	g_3 = -4 \left( 3a_\ast^2 + 2D_\ast^2  \right),
\;\;\;
	g_4 = 9 + 4a_\ast^2 +2D_\ast^2,
\;\;\;
	g_5 = -6,
\;\;\;
	g_6 = 1
\end{gather*}
and
\begin{gather*}
	\bar{e}_0 = D_\ast^4,
\;\;\;
	\bar{e}_1 = -4 D_\ast^2,
\;\;\;
	\bar{e}_2 = -2D_\ast^2 +8D_\ast,
\;\;\;
	\bar{e}_3 = -4,
\;\;\;
	\bar{e}_4 = 1,
\\
	\bar{g}_0 = D_\ast^4,
\;\;\;
	\bar{g}_1 = -4D_\ast^2,
\;\;\;
	\bar{g}_2 = 4+ 2D_\ast^2,
\;\;\;
	\bar{g}_3 = -4,
\;\;\;
	\bar{g}_4 = 1.
\end{gather*}

We define the following function:
\begin{eqnarray}
    \psi (r_\ast; a_\ast , i ) \coloneqq
    \begin{cases}
    \sin ^2 \alpha & (a_\ast \neq 0 \,\, \mathrm{and} \,\, i \neq 0) \\
    \infty & (a_\ast = 0 \,\, \mathrm{or} \,\, i = 0) \\
    \end{cases}
    .
\end{eqnarray}
That is, an element $(\psi (r_\ast; a_\ast , i ), \cos ^2 \beta(r_\ast; D_\ast , a_\ast )  ) \in \{ R(r_\ast) \cup \infty \} \times R(r_\ast)$ represents the apparent shape at a given parameter, where $R(r_\ast)$ is the field of rational functions. The element $(\infty , \cos^2\beta (r_\ast ; D_\ast , 0))$ corresponds to the apparent shape of the Schwarzschild black hole. The element $(\infty , \cos^2\beta (r_\ast ; D_\ast , a_\ast))$ ($a_\ast \neq 0$) corresponds to the apparent shape of the Kerr black hole as seen by an observer on the rotational axis.

From here on, note that two same irreducible rational functions have same coefficients. A reducible rational function does not always so.

For the case of a nonextreme Kerr black hole, we assume that $\psi (r_\ast ; a_\ast, i) = \psi (r_\ast ; a_\ast ^\prime, i ')$. From the coefficient $b_4$, we obtain $a_\ast ^2 \sin ^2 i = a_\ast ^{\prime 2} \sin ^2 i^\prime$. Then, analyzing the coefficient $a_3$, we conclude that $a_\ast = a_\ast ^\prime$ and $i = i ^\prime$. Additionally, we assume that $\cos ^2 \beta (r_\ast; D_\ast , a_\ast) = \cos ^2 \beta (r_\ast; D_\ast ^\prime, a_\ast ^\prime )$. From the coefficients $e_0$ and $e_3$, we deduce that $D_\ast = D_\ast^\prime$ and $a_\ast = a_\ast^\prime$, respectively.

Now, considering the case of an extreme Kerr black hole, we assume that $\psi (r_\ast ; a_\ast, i) = \psi (r_\ast ; a_\ast ^\prime, i ')$. By comparing $\bar{a}_2$, we can determine that $i = i^\prime$. After all, this assumption leads to $a\ast = a_\ast^\prime$ and $i = i^\prime$. We also assume that $\cos ^2 \beta (r_\ast; D_\ast , a_\ast) = \cos ^2 \beta (r_\ast; D_\ast ^\prime, a_\ast ^\prime )$. Comparing $\bar{e}_0$, we can establish that $D\ast = D_\ast^\prime$. Consequently, the assumption implies that $D_\ast = D_\ast^\prime$ and $a_\ast = a_\ast^\prime$.

Since the transformation from celestial coordinates $(\alpha, \beta)$ to screen coordinates $(x, y)$ is injective, the map $\tilde{f}_1 : P/\sim \rightarrow S$ is proved to be injective, which means that an apparent shape of a bare Kerr black hole is unique.

\section{Parameter determination based on features of apparent shape.}
\label{sec:pc}

\subsection{Observables for apparent shape}
\label{subsec:pc1}
\begin{figure}[tb]
		\begin{tabular}{ c }
			\includegraphics[height=3.8cm]
            {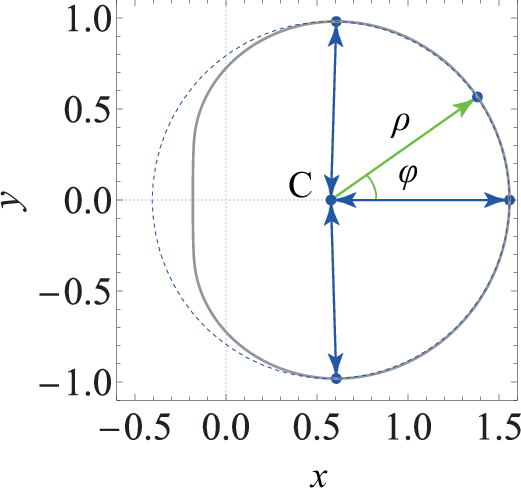}
		\end{tabular}
	\caption{\footnotesize{Polar coordinates $(f,s)$ with origin at the center C of the apparent shape of the Kerr black hole.}}
	\label{fig030}
\end{figure}
\begin{figure}[tb]
		\begin{tabular}{ c }
			\includegraphics[height=5.5cm]
            {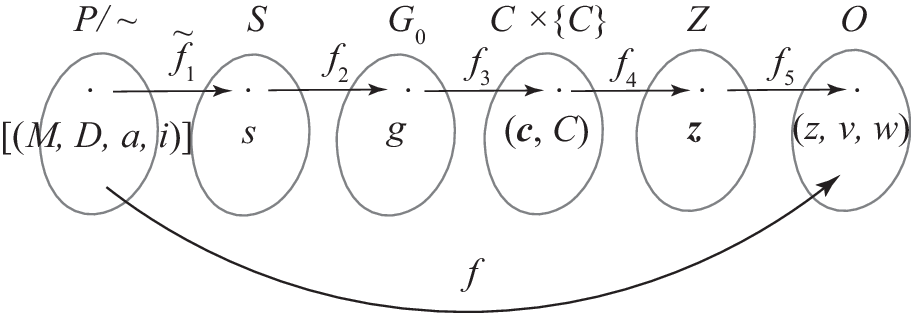}
		\end{tabular}
	\caption{\footnotesize{Composite map $f$ from the quotient parameter space $P/\sim$ to the observable space $O$}}
	\label{fig040}
\end{figure}
\begin{figure}[tb]
		\begin{tabular}{ ccc }
			\includegraphics[height=4.3cm]{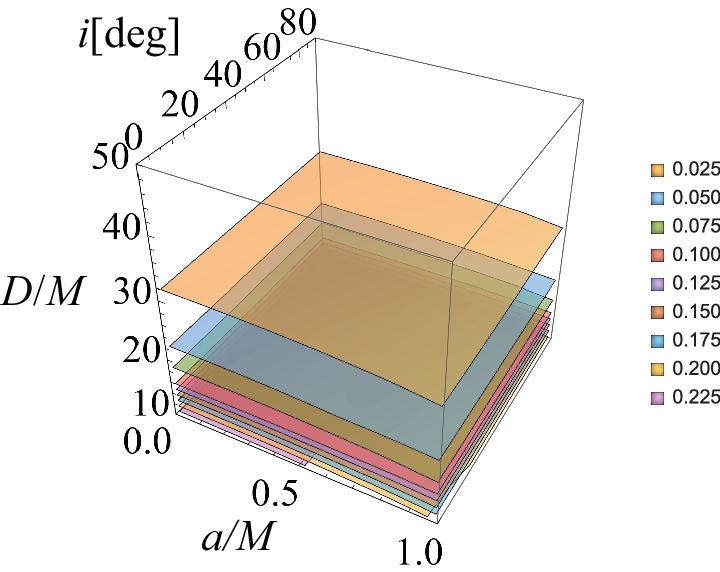} &
			\includegraphics[height=4.3cm]{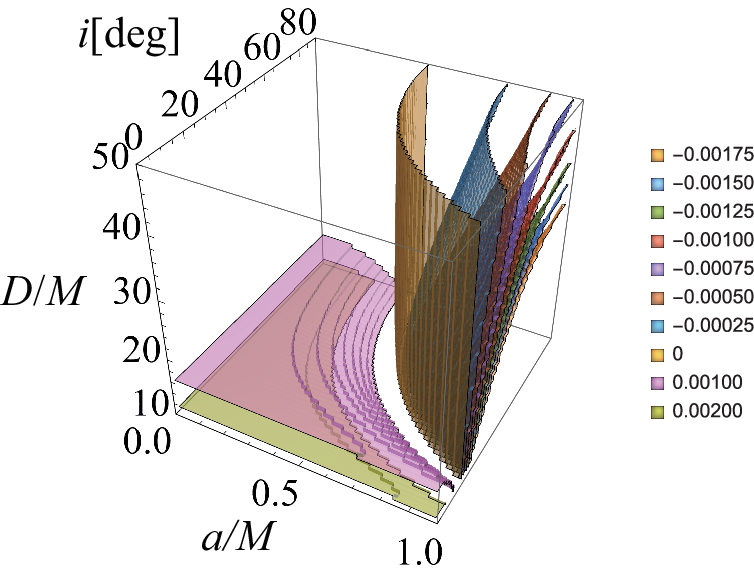} &
			\includegraphics[height=4.3cm]{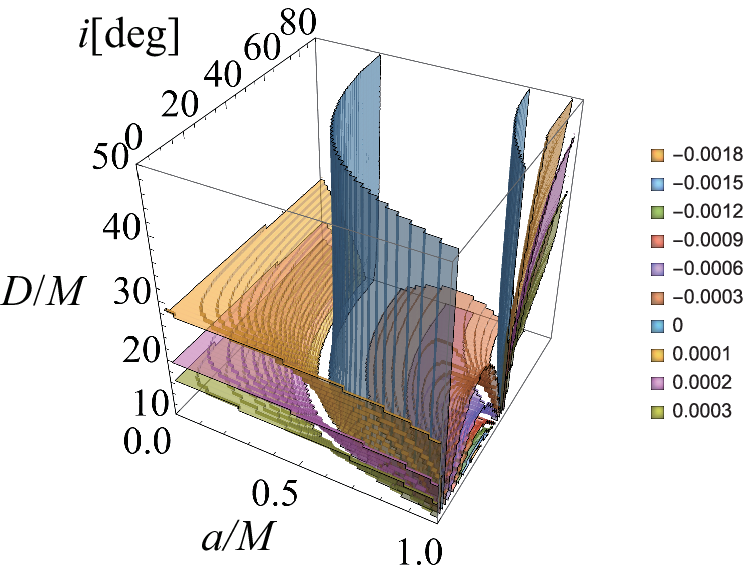} \\
			(a) $z$ & (b) $v$ & (c) $w$ \\
		\end{tabular}
	\caption{\footnotesize{(a) A contour map of the size $z$. (b) A contour map of the primary distortion $v$. (c) A contour map of the secondary distortion $w$.}}
	\label{fig050}
\end{figure}
We would like to propose observables of apparent shape to distinguish between all apparent shapes. An observable is defined as a quantity that can be derived from an apparent shape to describe its features. Since we explain the process of obtaining the observables here, readers who only want to know the observables obtained can go to the final paragraph of Sec.~\ref{subsec:pc1}.

When the observer is at spatial infinity, the radius and distortion parameter are useful to distinguish between the apparent shapes~\cite{Hioki:2009na}. However, when the distance of the black hole is finite, the same approach does not work as when the distance is spatial infinity, because the change in distance affects both size and shape of the apparent shape.

We assume a situation where we have no prior information on the location of the black hole. Observables, as defined, must be derivable from the apparent shape itself. As shown in Fig.~\ref{fig030}, we first define the center C of the apparent shape as the center of a circle passing through the top, rightmost, and bottom points of the apparent shape. In the case of the Schwarzschild black hole, this center is the point constituted by the principal null geodesic that connects the observer straight to the singularity. The positions of the points that constitute the apparent shape are then expressed in terms of radius $\rho$ and angle $\varphi$ with respect to this center C. 
Using $\rho$ and $\varphi$, we define a map $f_2$ from the apparent-shape library into a set of all simple closed curves $G$ as follows:
\begin{eqnarray}
	f_2 : S \ni s \mapsto \{ \left( \rho(\varphi)\cos \varphi, \rho(\varphi)\sin \varphi \right) | 0 \leq \varphi < 2\pi \} \in G \, .
\end{eqnarray}
Let us denote the range of the map $f_2$ by $G_0= f_2(S) $. We call $G_0$ a centered apparent-shape library and its element a centered apparent shape.

We will now consider the complex Fourier series of the function $\rho(\varphi)$.
\begin{eqnarray}
	\rho(\varphi) = \sum_{n = - \infty}^{\infty}c_n e^{in\varphi},
\;\;\;
	c_n = \frac{1}{2\pi}\int_{-\pi}^{\pi} \rho(\varphi) e^{-in\varphi}{\rm d}\varphi.
\end{eqnarray}
Note that $\rho(\varphi)$ is a real function, so $c_{-n} = c_{n}^{*}$ holds. Since the function $\rho(\varphi)$ is piecewise smooth, the complex Fourier series of the function $\rho(\varphi)$ converges to the function. Since the apparent shape has symmetry of $\rho(-\varphi)=-\rho(\varphi)$, the Fourier coefficients $c_n$ are real.

We use up to the eighth Fourier coefficients in our analysis. Let us denote by $d$ the map from a centered apparent shape to eight Fourier coefficients ${\bm c} := \left( c_0, c_1, c_2, c_3, c_4, c_5, c_6, c_7 \right)$.

We need to define the following map $f_3$ so that we can perform principal component analysis of the Fourier coefficients:
\begin{eqnarray}
	f_3 : G_0 \ni g \mapsto ( d(g), C ) \in C \times \{ C \} \, ,
\end{eqnarray}
where Fourier coefficient space $C$ is defined as $C \coloneqq d \left( G_0 \right) \subset \mathbb{R}^8$. Principal component analysis is a method to derive new variables, called the principal components, from existing variables in such a way that maximizes the variance in given data. (See Appendix~\ref{sec:pca} for a general review on the principal component analysis.) We perform the principal component analysis in $C$ to construct the observables as the principal components that are capable of distinguishing all apparent shapes.

We consider a map $f_4$ from the Fourier coefficients to the principal components. The principal component space $Z$ is the set of orthogonally transformed elements of the Fourier coefficient space $C$. This map is defined as follows:
\begin{eqnarray}
	f_4 : C \times \{ C \} \ni ({\bm c}, C) \mapsto {\bm z} := (z_1, z_2, z_3, z_4, z_5, z_6, z_7, z_8) \in Z \, .
\end{eqnarray}
It should be noted here that the domain of the map $f_4$ is not $C$ but $C \times \{C\}$. Since principal component analysis calculates the variance of the entire dataset, principal component analysis requires $\{ C \}$ as input.

In general, a map from multiple variables to principal components is an orthogonal transformation. The map $f_4 (\cdot , C)$ is an orthonormal transformation. We present the transformation matrix $\bm{A}$ of the orthonormal transformation in Appendix~\ref{sec:matrix}. To find $\bm{A}$, $50^3$ elements are sampled from $C$. We provide the sampled data as Supplemental Material~\cite{SupplmentalMaterial} to the paper.

A set of all elements of the first three components of the principal component space $Z$ is defined as an observable space $O$. We define projection $f_5$ as follows:
\begin{eqnarray}
	f_5 : Z \ni {\bm z} \mapsto (z_1, z_2, z_3) \in O \, .
\end{eqnarray}

We define the composite map $f$ (see Fig.~\ref{fig040}) from the quotient space $P/\sim$ to the observables space $O$ as follows:
\begin{eqnarray}
	f := f_5 \circ f_4 \circ f_3 \circ f_2 \circ \tilde{f}_1 \, .
\end{eqnarray}
The first, second, and third components of $(z_1, z_2, z_3)$ in $f \left( P/\sim \right)$ will be referred to as the size $z$, primary distortion $v$, and secondary distortion $w$, respectively. Figure~\ref{fig050} depicts the contour surfaces of the observables $z$, $v$, and $w$. In Sec.~\ref{sec:determination}, we investigate whether the maps $f$ is injective and show that the values of the parameters can be determined from the apparent shape.

In summary, we utilize three observables to characterize the apparent shape: size $z$, primary distortion $v$, and secondary distortion $w$. These observables are derived from the Fourier coefficients of the apparent shape. By orthogonal transformation of the Fourier coefficients up to the eighth, we obtain eight quantities. We select three quantities that mainly characterize the apparent shape. This orthogonal transformation is conducted using principal component analysis, a standard data analysis method that reduces dimensions in multivariable data. The orthogonally transformed quantities are referred to as principal components.

\subsection{Parameter determination}
\label{sec:determination}
\begin{figure}[tb]
		\begin{tabular}{ ccc }
			\includegraphics[height=4.3cm]{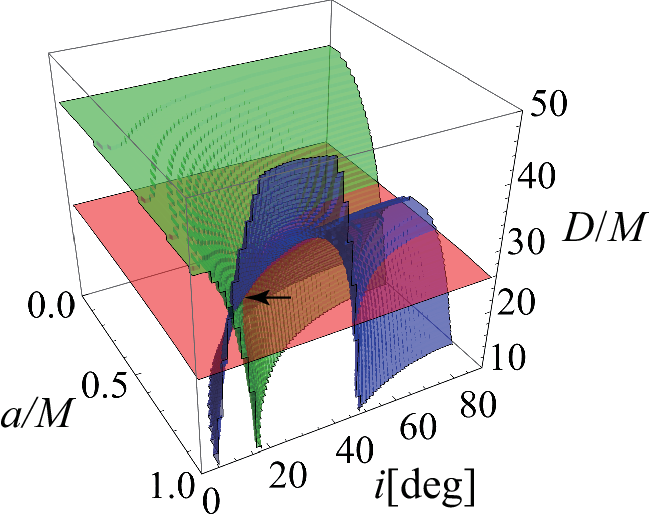} &
			\includegraphics[height=4.3cm]{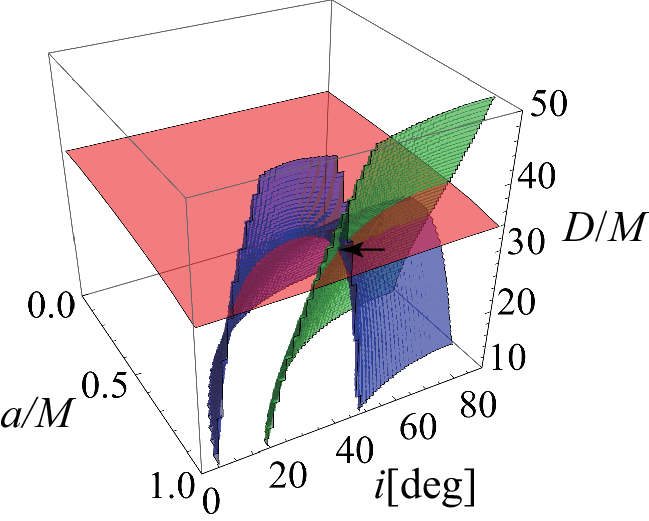} &
			\includegraphics[height=4.3cm]{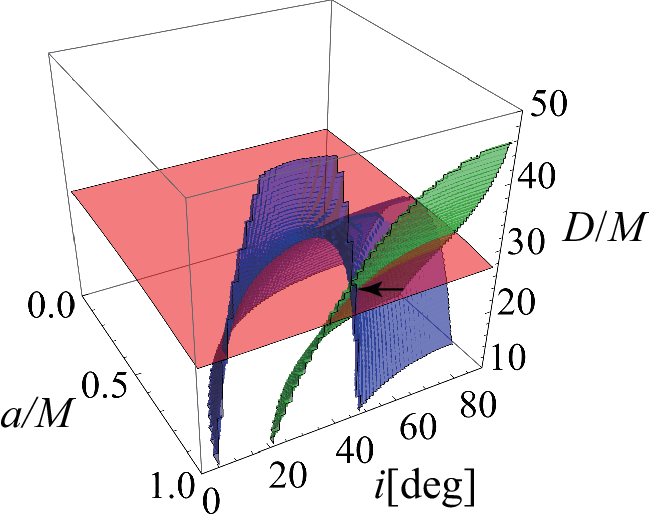} \\
			(a) $v=0.00015$, $w=-0.00006$ & (b) $v=-0.0005$, $w=-0.00006$ & (c) $v=-0.001$, $w=-0.00006$ \\
			\includegraphics[height=4.3cm]{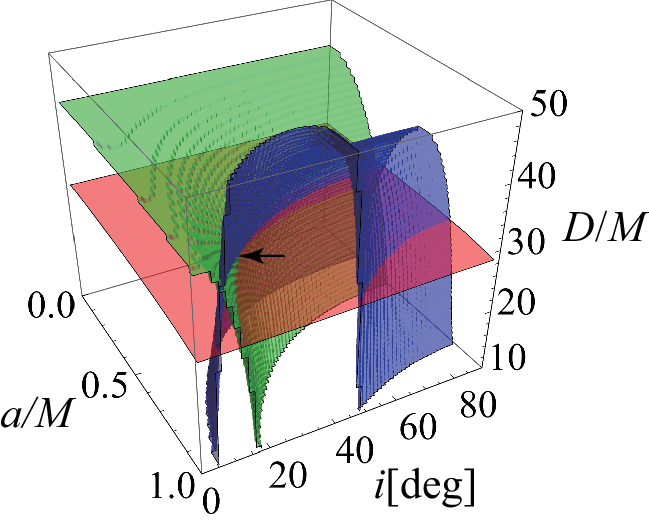} &
			\includegraphics[height=4.3cm]{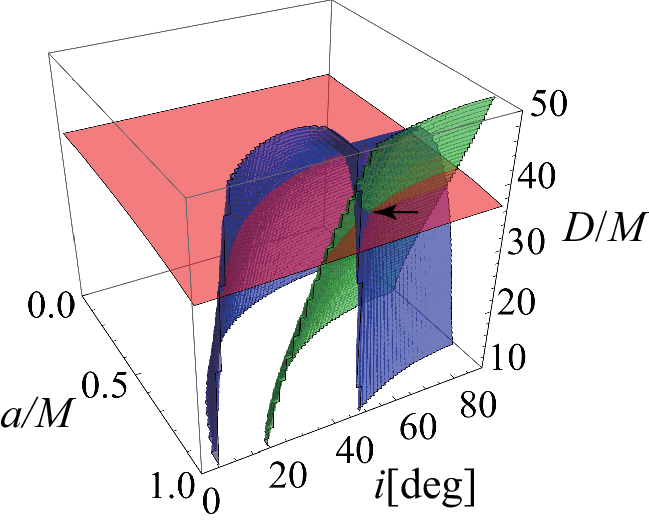} &
			\includegraphics[height=4.3cm]{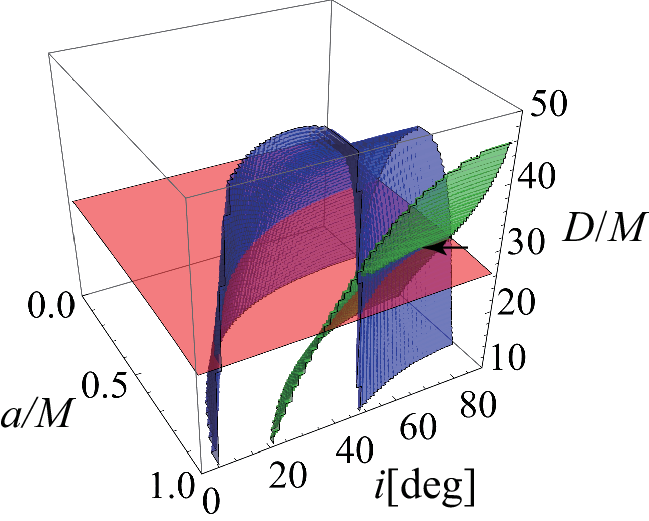} \\
			(d) $v=0.00015$, $w=-0.00003$ & (e) $v=-0.0005$, $w=-0.00003$ & (f) $v=-0.001$, $w=-0.00003$ \\
			\includegraphics[height=4.3cm]{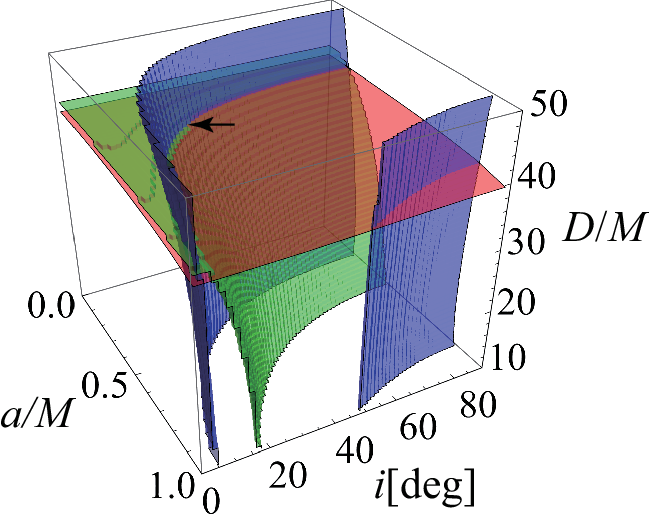} &
			\includegraphics[height=4.3cm]{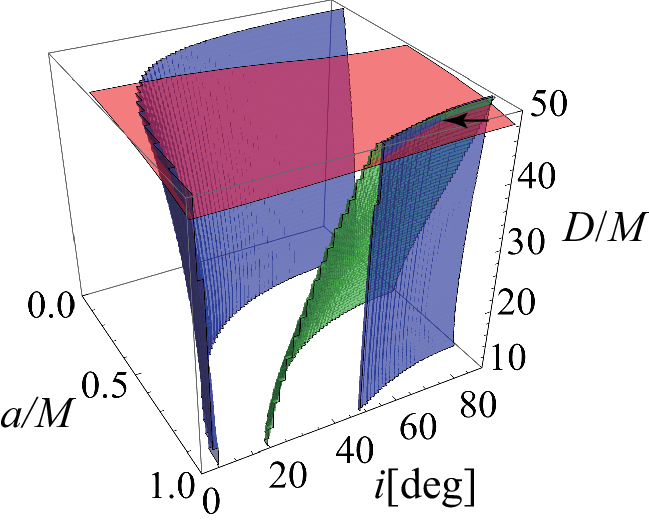} &
			\includegraphics[height=4.3cm]{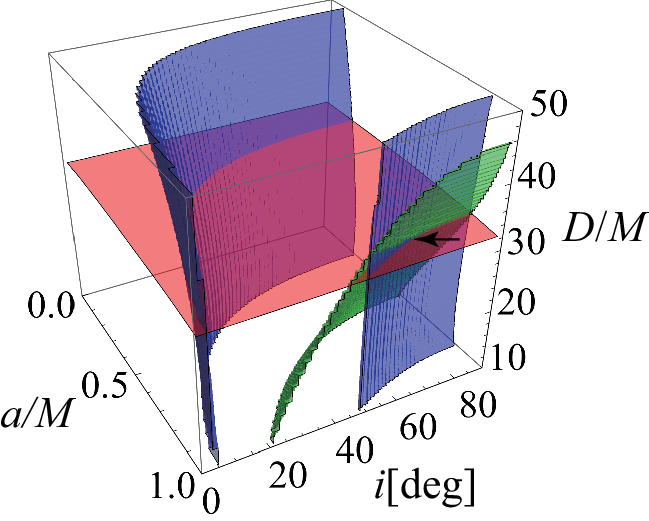} \\
			(g) $v=0.00015$, $w=0.00002$ & (h) $v=-0.0005$, $w=0.00002$ & (i) $v=-0.001$, $w=0.00002$ \\
		\end{tabular}
	\caption{\footnotesize{The contours of the size $z$ (the red sufaces), the primary distortion $v$ (the green surfaces), and the secondary distortion $w$ (the blue surfaces). In (a), (b), (c), (d), (e), (f), (g), (h), and (i), sizes $z$ are 0.035, 0.02, 0.03, 0.028, 0.017, 0.033, 0.0146, 0.0099, and, 0.022, respectively.}}
	\label{fig090}
\end{figure}
Since the apparent shape of the Kerr black hole is unique and the map $\tilde{f}_1$ is injective, in principle it is possible to determine the parameters from the apparent shape. We propose a method that can determine the parameters from the observables of the apparent shape.

In Fig.~\ref{fig050}(a), the size of the apparent shape becomes smaller as $D_\ast$ increases. In addition, the size of the two apparent shapes is slightly smaller at the same $D_\ast$ for the one with larger $a_\ast$. Figure~\ref{fig050}(b) shows that the primary distortion of the apparent shape reflects the contrast between the apparent shape when $a_\ast$ and $i$ are large and when either $a_\ast$ or $i$ is small. Looking at Fig.~\ref{fig050}(c), the secondary distortion of the apparent shape highlights the common distortion effect between the apparent shape when both $a_\ast$ and $i$ are large and when either $a_\ast$ or $i$ is small.

While varying the value of the observables $(z, v, w)$, we draw together the contours of the observables $(z, v, w)$ in Fig.~\ref{fig090}. It can be seen that in all cases the three contours intersect at a single point. This fact means that the map $f : P/\sim \rightarrow O$ is injective. Once the values of the observables $(z, v, w)$ is obtained from the shadow observations, the parameters $(D/M, a/M, i)$ can be determined.

\subsection{Demonstration}
\label{sec:demo}
In this section, we will demonstrate the use of our method. Consider the case where one observes the shadow of a black hole. First, one computes the Fourier coefficients $\bm{c}$ up to the eighth of the apparent shape of the black hole. Suppose the Fourier coefficients have the following value:
\begin{eqnarray}
	\bm{c}^\prime = \begin{pmatrix} 0.033407 \\ 0.0011385 \\ -0.0022276 \\ -0.00013489 \\ -0.0014535 \\ -0.00028273 \\ -0.0014040 \\ -0.00027413  \end{pmatrix},
\end{eqnarray}
where the prime denotes the transpose of a matrix. Then, one uses the transformation matrix $\bm{A}$ presented in Appendix~\ref{sec:matrix} to find the principal component $\bm{z}$ as follows:
\begin{eqnarray}
	\bm{z}^\prime = \bm{A}^\prime \bm{c}^\prime = \begin{pmatrix} 0.033546\\ -0.0010929 \\ -3.0965 \times 10^{-5} \\ -5.6672 \times 10^{-5} \\ 8.5769 \times 10^{-6}\\ -1.2802  \times 10^{-6}\\ -2.5584  \times 10^{-7}\\ 6.3480 \times 10^{-7} \end{pmatrix} \, .
\end{eqnarray}
The first three principal components of ${\bm z}'$ are the observables $z$, $v$, and $w$. We have included data on the parameters and their corresponding principal components in the Supplemental Material~\cite{SupplmentalMaterial}. With the data in the Supplemental Material, one can plot contours of principal component $z,$ $v$, and $w$ in $(D_\ast, a_\ast, i)$ space. The contours in three values intersect at a single point in $(D_\ast, a_\ast, i)$ space. Finally, the values of the parameters can be determined as follows:
\begin{eqnarray}
	(D/M, a/M, i) = (26.3, 0.81, 72.9) \, .
\end{eqnarray}

\section{Discussion}
\label{sec:discussion}
We have shown that the dimensionless parameters $(D/M, a/M ,i)$ of a black hole can be determined only by observing the apparent shape of the shadow. Then, is it possible to determine the four parameters $(M, D, a, i)$ from the apparent shape of the shadow? Unfortunately, the four parameters cannot be determined from the observation of the apparent shape alone, as can be seen from the fact that the apparent shape is determined by the dimensionless parameters $(D/M, a/M, i)$.

If one of the values of $(M, D, a)$ is known {\it a priori} from another observation, then $(M, D, a, i)$ can be uniquely determined. In terms of accompanying the observation of the apparent shape of the shadow, another candidate would be to observe the flux of the accretion flow or accretion disk, which is the light source. In other words, if we have an image of both the bright ring and the shadow, we can determine all four parameters $(M, D, a, i)$.

The method to determine the parameters including the mass was constructed targeting the Schwarzschild black hole with a thin accretion disk by the present authors~\cite{Hioki:2022mdg}. The key point for determining the mass is that the flux of the accretion disk is a function of the mass.

In the present paper, we assume a bare black hole, and therefore there is no accretion flow or accretion disk. Since the subject is whether the parameters can be determined from the apparent shape of the shadow only, we do not consider information from other observations

We obtained the principal components based on Fourier coefficients up to the eighth order. In the process to develop our method, we initially attempted to use the Fourier coefficients up to the third order to see if they could be used directly as three observables. However, there was no one-to-one correspondence between the dimensionless parameters $(D/M, a/M, i)$ and the observables. Therefore, our approach was to sequentially increase the Fourier coefficients up to the order at which a one-to-one correspondence is found. When considering Fourier coefficients of fourth order or higher, we did not use the Fourier coefficients themselves, but rather obtained the principal components, because we wanted to maximize the variance of the coefficients and visualize them in a three-dimensional space.

Following this approach, we learned to know that the Fourier coefficients up to the eighth order are necessary, and the three principal components based on these coefficients can be used as the observables. Of course, if the Fourier coefficients up to ninth or higher orders are included, a one-to-one correspondence between the principal components and the dimensionless parameters $(D/M, a/M, i)$ will be obtained.

In this paper, three of the eight principal components with the largest variance were employed as observables. The remaining five principal components are also quantities that represent the distortion of the apparent shape from the circle. It is possible to choose any three among these quantities and look for a one-to-one correspondence with the parameters. Since our goal is to find a set of observables that have a one-to-one correspondence with the parameters, the finding of one combination of principal components is sufficient.

In the future, if we capture a high-resolution image of a black hole, we will likely observe a higher order ring. How many principal components will be needed to analyze the higher-order ring? Let us consider a model of a black hole. The key point is how many parameters the model has. For example, if there are parameters such as the mass, distance, angular momentum, inclination angle, and cosmological constant, then four principal components, the number of parameters five minus one, are needed as observables. This expectation of the number of observables assumes that there is a one-to-one correspondence between observables and dimensionless parameters.

When observing a higher-order ring, an important question is whether an injection or a one-to-one correspondence exists between the parameters of a black hole and observables. This will be one of the challenging themes for the future research on shadow observation.

Let us consider applying our method to the images of M87 and ${\rm Sgr~A}^\ast$ obtained by the Event Horizon Telescope. Although the current Event Horizon Telescope image shows the bright rings, the shadow, which is the dim part of the image, is not fully identified~\cite{EventHorizonTelescope:2019ths}. It means that we know the rough size of the radius of the shadow, but not the distortion of the shadow~\cite{EventHorizonTelescope:2021dqv, Vagnozzi:2022moj}. Since this is a problem of observation accuracy, improvements in observation equipment are underway. If such improvements are made and the shadows of M87 and ${\rm Sgr~A}^\ast$ can be observed with high accuracy, then the apparent shapes can be obtained from the image.

If the shadows of M87 and ${\rm Sgr~A}^\ast$ are identified, then the observed data can be used to validate our method. We can perform the determination of the parameters of M87 and ${\rm Sgr~A}^\ast$ using the method.

This method will be essential in the future when there is an opportunity to observe a black hole from a close distance or we need to determine the parameters with necessary accuracy. It is important to continue to study shadows from the perspective of preparing for such times.

\section{Conclusion}
\label{sec:conclusion}
We have analyzed the apparent shape of Kerr black holes without accretion disks, i.e., bare Kerr black holes. We have investigated the determination of parameters from the apparent shape, assuming that the black hole's distance is finite. The parameters of the system consisting of the black hole, the light source, and the observer are the mass $M$, the specific angular momentum $a$, the inclination angle $i$, and the distance $D$.

We have defined observables for the apparent shape and proposed a new method to determine the parameters of a black hole by shadow observations. In this method, one first performs a Fourier analysis of the observed apparent shape to obtain the Fourier coefficients up to the eighth order. The vector of Fourier coefficients is then transformed by the orthogonal matrix we have presented. The first three principal components obtained are the values of the observables $(z, v, w)$. From their values and Fig.~\ref{fig050}, one could determine the values of the parameters $(D/M, a/M, i)$.

We also proved the uniqueness of the apparent shape of the bare Kerr black hole. This uniqueness reveals the structure of the apparent shape and image libraries, which is found for the first time in this paper.

Our method can be used for shadows of various black hole solutions and for shadows of black holes in various situations. We considered the Kerr black hole without the accretion disk for simplicity. The next step would be to consider more realistic models with accretion disks. The reversibility of the map from the parameter space to the image library should be examined in such realistic models. If the image library is extended to include various black hole solutions as models, and if a shadow image not included in such an image library is actually observed, it suggests the existence of a new black hole solution or the modification of the gravitational theory. Finally, constructing a movie library involving time-varying shadow images in preparation for future observations would be challenging but interesting.

\section*{Acknowledgements}
We thank the anonymous referee for very helpful comments. This work was supported in part by JSPS KAKENHI Grant No. JP22K03623 (U.M.).

\appendix

\section{Principal component analysis}
\label{sec:pca}
Principal component analysis is a technique to analyze large datasets with high number of dimensions with enhancing data interpretability, retaining maximum information, and enabling multidimensional data visualization (for example, see~\cite{Jolliffe} and references therein).

Let us consider $P(\geq 1)$-dimensional data of $N (\geq 1)$ samples. In our case, $N$ corresponds to the number of points in the space of $(D/M, a/M,i)$. On the other hand, $P$ corresponds to the number of Fourier components. The value of $j$th variate of $i$th sample is denoted by $x_{ij} \in {\mathbb R} \; ( 1\leq  i \leq N, 1 \leq j \leq P)$. The mean and covariance are defined by
\begin{align}
	\bar{x}_j
	:=
	\frac{1}{N} \sum_{i=1}^N x_{ij},
\;\;\;\
	\Cov(x_j,x_k)
	:=
	\frac{1}{N-1}\sum_{i=1}^N (x_{ij}-\bar{x}_j)(x_{ik}-\bar{x}_k).
\label{eq:x-mean-cov}
\end{align}
We consider the following transformation from the original data $x$ to principal components $z$,
\begin{align}
	z_{ij}
	=
	\sum_{k=1}^P a_{jk} x_{ik}
\;\;\;
	(1 \leq i \leq N; 1 \leq j \leq P),
\label{eq:z-x-relation}
\end{align}
where $a_{jk}$ is a constant to be determined later. The mean and covariance of principal components are represented by those of original data and $a_{jk}$,
\begin{align}
	\bar{z}_j
	&:=
	\frac{1}{N} \sum_{i=1}^N z_{ij}
	=
	\sum_{k=1}^P a_{jk} \bar{x}_k,
\\
	\Cov(z_j, z_k)
	&:=
	\frac{1}{N-1}\sum_{i=1}^N (z_{ij}-\bar{z}_{j})(z_{ik}-\bar{z}_{k})
	=
	\sum_{l=1}^P\sum_{m=1}^P a_{jl}a_{km} \Cov(x_{l},x_{m}).
\label{eq:z-mean-cov}
\end{align}

For simplicity, we introduce the following matrices and vectors,
\begin{align}
\begin{split}
	{\bm X} &:= ({\bm x}_1 \; {\bm x}_2 \; \cdots \; {\bm x}_P),
\;\;\;
	{\bm x}_j := (x_{1j} \; x_{2j} \; \cdots \; x_{Nj})',
\\
	{\bm Z} &:= ({\bm z}_1 \; {\bm z}_2 \; \cdots \; {\bm z}_P),
\;\;\;
	{\bm z}_j := (z_{1j} \; z_{2j} \; \cdots \; z_{Nj})',
\\
	{\bm A} &:= ({\bm a}_1 \; {\bm a}_2 \; \cdots \; {\bm a}_P),
\;\;\;
	{\bm a}_j
	:=
	( a_{j1} \; a_{j2} \; \cdots \; a_{jP} )',
\\
	{\bm C} &:= (C_{jk}) := ( \Cov(x_j, x_k) ),
\end{split}
\label{eq:mat-vec-notation}
\end{align}
where $1 \leq j \leq P, 1 \leq k \leq P$ and the prime denotes the transpose of a matrix. We call ${\bm A}$ the $transformation$ $matrix$. If one adopts the above notations, Eqs.~\eqref{eq:z-x-relation} and \eqref{eq:z-mean-cov} are written as
\begin{gather}
	{\bm z}_j
	=
	{\bm X}{\bm a}_j,
\;\;\;
	{\bm Z}
	=
	{\bm X}{\bm A},
\label{eq:zXa-ZXA}
\\
	\Cov (z_k, z_j)
	=
	{\bm a}_k' {\bm C} {\bm a}_{j}.
\label{eq:aCa}
\end{gather}

For given data ${\bm X}$, we will find the first principal component ${\bm z}_1 = {\bm X}{\bm a}_1$ so that it maximizes variance $\Var(z_1) := \Cov (z_1,z_1) = {\bm a}_1' {\bm C} {\bm a}_1$ under normalization condition ${\bm a}_1' {\bm a}_1=1$. We implement this by defining
\begin{align}
	F_1
	:=
	{\bm a}_1' {\bm C} {\bm a}_1 -\lambda_1 ( {\bm a}_1' {\bm a}_1-1 )
	=
	\sum_{j=1}^P \sum_{k=1}^P C_{jk} a_{1j} a_{1k} 
	-
	\lambda_1 \left( \sum_{j=1}^P a_{1j}^2-1 \right),
\end{align}
where $\lambda_1$ is the Lagrange multiplier. Differentiating $F_1$ by ${\bm a}_1$ and demanding it to vanish, we obtain
\begin{align}
	\frac{ \pd F_1 }{\pd {\bm a}_1 }
	=
	2{\bm C}{\bm a}_1 -2\lambda_1 {\bm a}_1
	= {\bm 0}_P,
\end{align}
where ${\bm 0}_P$ is a $P$-dimensional zero vector. Thus, we find that ${\bm a}_1$ is an eigenvector of ${\bm C}$ belonging to eigenvalue $\lambda_1$:
\begin{align}
	{\bm C} {\bm a}_1
	=
	\lambda_1 {\bm a}_1.
\label{eq:Ca1}
\end{align}

In order to determine the $j$th principal component ${\bm z}_j = {\bm X} {\bm a}_j$ ($2 \leq j \leq P$), we define
\begin{align}
	F_j
	:=
	{\bm a}_j' {\bm C} {\bm a}_j -\lambda_j ( {\bm a}_j' {\bm a}_j-1 )
	- \sum_{k=1}^{j-1} \lambda_{jk}  {\bm a}_j' {\bm a}_k
\;\;\;
	(2 \leq j \leq P),
\end{align}
where $\lambda_j$ and $\lambda_{jk}$ are Lagrange multipliers. The last term corresponds to the orthogonality between ${\bm a}_j$ and ${\bm a}_k \; (1 \leq k \leq j-1)$. Demanding $\frac{ \pd F_j }{\pd {\bm a}_j } = {\bm 0}_P$ and using the proof by mathematical induction, we can show that $\lambda_{jk}=0$ and ${\bm C} {\bm a}_j = \lambda_j {\bm a}_j \; (2 \leq j \leq P, 1 \leq k \leq j-1)$ hold as follows: From $\frac{ \pd F_j }{\pd {\bm a}_j } = {\bm 0}_P$, we obtain
\begin{align}
	2{\bm C}{\bm a}_j -2\lambda_j {\bm a}_j - \sum_{k=1}^{j-1} \lambda_{jk}{\bm a}_k
	=
	{\bm 0}_P
\;\;\;
	(2 \leq j \leq P).
\label{eq:Fj_derivative}
\end{align}
Setting $j=2$ in Eq.~\eqref{eq:Fj_derivative} and multiplying it by ${\bm a}_1'$, we find that $\lambda_{21}=0$ and ${\bm C}{\bm a}_2 = \lambda_2 {\bm a}_2$ hold using Eq.\ \eqref{eq:Ca1} $(\Leftrightarrow {\bm a}_1' {\bm C} = \lambda_1 {\bm a}_1')$ and ${\bm a}_1' {\bm a}_2=0$. Now, suppose that $ \lambda_{kl}=0 $ and ${\bm C}{\bm a}_{k} = \lambda_{k} {\bm a}_{k} \; (1 \leq j \leq P-1; 2 \leq k \leq j, 1 \leq l \leq k-1)$ hold. Setting $j \to j+1$ in Eq.\ \eqref{eq:Fj_derivative} and multiplying it by ${\bm a}_m' \; (1 \leq m \leq j)$, we obtain $\lambda_{j+1, m}=0 \; (1 \leq m \leq j)$ and ${\bm C}{\bm a}_{j+1} = \lambda_{j+1} {\bm a}_{j+1}$. 
Thus, we have shown
\begin{align}
	{\bm C}{\bm a}_j = \lambda_i {\bm a}_j,
\;\;\;
	{\bm a}_j' {\bm a}_k = \delta_{jk}
\;\;\;
	(1 \leq j \leq P; 1 \leq k \leq P). 
\label{eq:eigenproblem2}
\end{align}

In summary, for a given dataset ${\bm X}$, one can obtain the principal components ${\bm Z}$ through Eq.~\eqref{eq:zXa-ZXA}. Here, ${\bm a}_j$ is given as the eigenvector of ${\bm C}$ belonging to eigenvalue $\lambda_j \; (\lambda_1 \geq \lambda_2 \geq \cdots \geq \lambda_P)$. Using Eqs.~\eqref{eq:aCa} and \eqref{eq:eigenproblem2}, one can see that the principal components have no correlations:
\begin{align}
	\Cov (z_k, z_j)
	=
	{\bm a}_k' {\bm C} {\bm a}_{j}
	=
	{\bm a}_k' \lambda_{j} {\bm a}_{j}
	=
	\lambda_{j} \delta_{jk}
	\;\;\;
	(1 \leq j \leq P; 1 \leq k \leq P).
\end{align}

\section{Matrix representation of orthogonal transformation}
\label{sec:matrix}
To find the transformation matrix ${\bm A}$ of the orthogonal transformation $f_4 (\cdot , C)$, $50^3$ elements as a dataset $C_s$ are sampled from $C$. The dataset is defined as
\begin{align}
C_s \coloneqq f_1\left( \{ 1 \} \times \{ 10, 14, \cdot \cdot \cdot , 50 \} \times \{ 0.01, 0.108, \cdot \cdot \cdot , 0.99 \} \times \{ 0.01, 0.16508, \cdot \cdot \cdot , 1.5608 \} \right) \, .
\end{align}
The matrix ${\bm C}$ can be obtained by computing the variance and covariance for $C_s$:
{\fontsize{6pt}{6pt}\selectfont
$$
\begin{pmatrix}
0.0024721 & 0.000019331 & -0.00011978 & -5.7135\times 10^{-6} & -0.00010400 & -8.4192\times 10^{-6} & -0.00010190&-8.1935\times 10^{-6}\\
0.000019331&1.1224\times 10^{-6} & -1.6612 \times 10^{-6}& -6.5713 \times 10^{-8}& -9.1028\times 10^{-7} & -2.5095 \times 10^{-7}&-8.4326 \times 10^{-7}&-2.5376 \times 10^{-7}\\
-0.00011978&-1.6612\times 10^{-6}&6.3453\times 10^{-6}&2.9388\times 10^{-7}&5.1117\times 10^{-6}&5.4764\times 10^{-7}&4.9720\times 10^{-6}&5.4019\times 10^{-7}\\
-5.7135\times 10^{-6}&-6.5713\times 10^{-8}&2.9388\times 10^{-7}&3.0433\times 10^{-8}&2.3731\times 10^{-7}&3.6572\times 10^{-8}&2.3363\times 10^{-7}&3.4675\times 10^{-8}\\
-0.00010400& -9.1028\times 10^{-7}&5.1117\times 10^{-6}& 2.3731\times 10^{-7}& 4.3869\times 10^{-6}& 3.6893\times 10^{-7}&4.2926\times 10^{-6}& 3.6048\times 10^{-7}\\
-8.4192\times 10^{-6}& -2.5095\times 10^{-7}& 5.4764\times 10^{-7}& 3.6572\times 10^{-8}&3.6893\times 10^{-7}& 7.45857\times 10^{-8}& 3.5366\times 10^{-7}& 7.3622\times 10^{-8}\\
-0.00010190& -8.4326\times 10^{-7}&  4.97200\times 10^{-6}&2.3363\times 10^{-7}&  4.2926\times 10^{-6}& 3.5366\times 10^{-7}& 4.2032\times 10^{-6}& 3.4488\times 10^{-7}\\
-8.1935\times 10^{-6}&-2.5376\times 10^{-7}&  5.4019\times 10^{-7}&3.4675\times 10^{-8}&  3.6048\times 10^{-7}& 7.3622\times 10^{-8}&  3.4488\times 10^{-7}& 7.2868\times 10^{-8}
\end{pmatrix}.
$$
}
The eigenvalues $\lambda _j$ $(1\leq j \leq 8)$ of matrix ${\bm C}$ are as follows:
\begin{align}
\begin{pmatrix}
\lambda _1 \\
\lambda _2 \\
\lambda _3 \\
\lambda _4 \\
\lambda _5 \\
\lambda _6 \\
\lambda _7 \\
\lambda _8
\end{pmatrix}
=
\begin{pmatrix}
0.0024867 \\
1.5959 \times 10^{-6}\\
3.6429 \times 10^{-8}\\
1.6382 \times 10^{-9}\\
2.4964 \times 10^{-10}\\
3.7214 \times 10^{-12}\\
1.2963 \times 10^{-13}\\
7.5729 \times 10^{-14}
\end{pmatrix}
.
\end{align}
We write down its transformation matrix ${\bm A}$ of the orthogonal transformation $f_4 (\cdot , C)$:
{\fontsize{10pt}{8pt}\selectfont
$$
\begin{pmatrix}
0.99705&0.039961&-0.012705&-0.053784&-0.034029&-0.0078209&0.0015676&0.00010205\\
0.0078164&-0.77935&0.14520&-0.49563&0.068744&-0.11356&-0.028452&0.32769\\
-0.048323&0.58053&-0.043101&-0.45271&0.029961&-0.14968&-0.083962&0.65075\\
-0.0023046&0.019550&0.67230&0.17503&-0.62811&0.20519&-0.20213&0.19872\\
-0.041946&0.075764&-0.21761&-0.57373&-0.26095&0.37678&-0.45644&-0.44447\\
-0.0033993&0.15090&0.50911&-0.16041&0.26799&-0.61444&-0.28370&-0.40302\\
-0.041099&0.035309&-0.12390&-0.25265&-0.59445&-0.41897&0.58658&-0.21184\\
-0.0033084&0.15444&0.45054&-0.31163&0.32465&0.47695&0.56423&-0.15742
\end{pmatrix}.
$$
}




\begin{thebibliography}{99}

\bibitem{EventHorizonTelescope:2019dse}
K.~Akiyama \textit{et al.} [Event Horizon Telescope],
Astrophys. J. Lett. \textbf{875}, L1 (2019)
doi:10.3847/2041-8213/ab0ec7
[arXiv:1906.11238 [astro-ph.GA]].

\bibitem{EventHorizonTelescope:2019uob}
K.~Akiyama \textit{et al.} [Event Horizon Telescope],
Astrophys. J. Lett. \textbf{875}, no.1, L2 (2019)
doi:10.3847/2041-8213/ab0c96
[arXiv:1906.11239 [astro-ph.IM]].

\bibitem{EventHorizonTelescope:2019jan}
K.~Akiyama \textit{et al.} [Event Horizon Telescope],
Astrophys. J. Lett. \textbf{875}, no.1, L3 (2019)
doi:10.3847/2041-8213/ab0c57
[arXiv:1906.11240 [astro-ph.GA]].

\bibitem{EventHorizonTelescope:2019ths}
K.~Akiyama \textit{et al.} [Event Horizon Telescope],
Astrophys. J. Lett. \textbf{875}, no.1, L4 (2019)
doi:10.3847/2041-8213/ab0e85
[arXiv:1906.11241 [astro-ph.GA]].

\bibitem{EventHorizonTelescope:2019ggy}
K.~Akiyama \textit{et al.} [Event Horizon Telescope],
Astrophys. J. Lett. \textbf{875}, no.1, L6 (2019)
doi:10.3847/2041-8213/ab1141
[arXiv:1906.11243 [astro-ph.GA]].

\bibitem{EventHorizonTelescope:2021bee}
K.~Akiyama \textit{et al.} [Event Horizon Telescope],
Astrophys. J. Lett. \textbf{910}, no.1, L12 (2021)
doi:10.3847/2041-8213/abe71d
[arXiv:2105.01169 [astro-ph.HE]].

\bibitem{EventHorizonTelescope:2021srq}
K.~Akiyama \textit{et al.} [Event Horizon Telescope],
Astrophys. J. Lett. \textbf{910}, no.1, L13 (2021)
doi:10.3847/2041-8213/abe4de
[arXiv:2105.01173 [astro-ph.HE]].

\bibitem{EventHorizonTelescope:2022xnr}
K.~Akiyama \textit{et al.} [Event Horizon Telescope],
Astrophys. J. Lett. \textbf{930}, no.2, L12 (2022)
doi:10.3847/2041-8213/ac6674.

\bibitem{EventHorizonTelescope:2022vjs}
K.~Akiyama \textit{et al.} [Event Horizon Telescope],
Astrophys. J. Lett. \textbf{930}, no.2, L13 (2022)
doi:10.3847/2041-8213/ac6675.

\bibitem{EventHorizonTelescope:2022wok}
K.~Akiyama \textit{et al.} [Event Horizon Telescope],
Astrophys. J. Lett. \textbf{930}, no.2, L14 (2022)
doi:10.3847/2041-8213/ac6429.

\bibitem{EventHorizonTelescope:2022exc}
K.~Akiyama \textit{et al.} [Event Horizon Telescope],
Astrophys. J. Lett. \textbf{930}, no.2, L15 (2022)
doi:10.3847/2041-8213/ac6736.

\bibitem{EventHorizonTelescope:2022urf}
K.~Akiyama \textit{et al.} [Event Horizon Telescope],
Astrophys. J. Lett. \textbf{930}, no.2, L16 (2022)
doi:10.3847/2041-8213/ac6672.

\bibitem{EventHorizonTelescope:2022xqj}
K.~Akiyama \textit{et al.} [Event Horizon Telescope],
Astrophys. J. Lett. \textbf{930}, no.2, L17 (2022)
doi:10.3847/2041-8213/ac6756.

\bibitem{darwin1959gravity}
C.~Darwin,
in ``The gravity field of a particle,''
Proceedings of the Royal Society of London. Series A.249, no.1257 (Jan., 1959) 180-194.

\bibitem{Bardeen:1973xx}
J.~M.~Bardeen,
in ``Black Holes,''
ed.\ C.\ DeWitt and B.\ DeWitt,
New York, Gordon \& Breach (1973).

\bibitem{Hioki:2008zw}
K.~Hioki and U.~Miyamoto,
Phys. Rev. D \textbf{78}, 044007 (2008),
[arXiv:0805.3146 [gr-qc]].

\bibitem{Bambi:2010hf}
C.~Bambi and N.~Yoshida,
Class. Quant. Grav. \textbf{27}, 205006 (2010)
doi:10.1088/0264-9381/27/20/205006
[arXiv:1004.3149 [gr-qc]].

\bibitem{Amarilla:2010zq}
L.~Amarilla, E.~F.~Eiroa and G.~Giribet,
Phys. Rev. D \textbf{81}, 124045 (2010)
doi:10.1103/PhysRevD.81.124045
[arXiv:1005.0607 [gr-qc]].

\bibitem{Amarilla:2013sj}
L.~Amarilla and E.~F.~Eiroa,
Phys. Rev. D \textbf{87}, no.4, 044057 (2013)
doi:10.1103/PhysRevD.87.044057
[arXiv:1301.0532 [gr-qc]].

\bibitem{Wei:2013kza}
S.~W.~Wei and Y.~X.~Liu,
JCAP \textbf{11}, 063 (2013)
doi:10.1088/1475-7516/2013/11/063
[arXiv:1311.4251 [gr-qc]].

\bibitem{Papnoi:2014aaa}
U.~Papnoi, F.~Atamurotov, S.~G.~Ghosh and B.~Ahmedov,
Phys. Rev. D \textbf{90}, no.2, 024073 (2014)
doi:10.1103/PhysRevD.90.024073
[arXiv:1407.0834 [gr-qc]].

\bibitem{Wei:2015dua}
S.~W.~Wei, P.~Cheng, Y.~Zhong and X.~N.~Zhou,
JCAP \textbf{08}, 004 (2015)
doi:10.1088/1475-7516/2015/08/004
[arXiv:1501.06298 [gr-qc]].

\bibitem{Singh:2017vfr}
B.~P.~Singh and S.~G.~Ghosh,
Annals Phys. \textbf{395}, 127-137 (2018)
doi:10.1016/j.aop.2018.05.010
[arXiv:1707.07125 [gr-qc]].

\bibitem{Stuchlik:2019uvf}
Z.~Stuchl\'\i{}k and J.~Schee,
Eur. Phys. J. C \textbf{79}, no.1, 44 (2019)
doi:10.1140/epjc/s10052-019-6543-8.

\bibitem{Hioki:2009na}
K.~Hioki and K.~i.~Maeda,
Phys. Rev. D \textbf{80}, 024042 (2009)
doi:10.1103/PhysRevD.80.024042
[arXiv:0904.3575 [astro-ph.HE]].

\bibitem{Hioki:2022mdg}
K.~Hioki and U.~Miyamoto,
Phys. Rev. D \textbf{107}, no.4, 044042 (2023)
doi:10.1103/PhysRevD.107.044042
[arXiv:2210.02164 [gr-qc]].

\bibitem{Abdujabbarov:2015xqa}
A.~A.~Abdujabbarov, L.~Rezzolla and B.~J.~Ahmedov,
Mon. Not. Roy. Astron. Soc. \textbf{454}, no.3, 2423-2435 (2015)
doi:10.1093/mnras/stv2079
[arXiv:1503.09054 [gr-qc]].

\bibitem{Kerr} R.~P.~Kerr, Phys.\ Rev.\ Lett.\ {\bf 11}, 237 (1963).

\bibitem{Chandrasekhar:1985kt}
S.~Chandrasekhar,
``The mathematical theory of black holes'',
Oxford Univ.\ Press (1992).

\bibitem{Grenzebach:2014fha}
A.~Grenzebach, V.~Perlick and C.~L\"ammerzahl,
Phys. Rev. D \textbf{89}, no.12, 124004 (2014)
doi:10.1103/PhysRevD.89.124004
[arXiv:1403.5234 [gr-qc]].

\bibitem{Chang:2020lmg}
Z.~Chang and Q.~H.~Zhu,
Phys. Rev. D \textbf{102}, no.4, 044012 (2020)
doi:10.1103/PhysRevD.102.044012
[arXiv:2006.00685 [gr-qc]].

\bibitem{EventHorizonTelescope:2021dqv}
P.~Kocherlakota \textit{et al.} [Event Horizon Telescope],
Phys. Rev. D \textbf{103}, no.10, 104047 (2021)
doi:10.1103/PhysRevD.103.104047
[arXiv:2105.09343 [gr-qc]].

\bibitem{SupplmentalMaterial}
See the Supplemental Material at http://link.aps.org/supplemental/10.1103/PhysRevD.109.044030 for table data on the parameters, their corresponding Fourier coefficients, and principal components.

\bibitem{Vagnozzi:2022moj}
S.~Vagnozzi, R.~Roy, Y.~D.~Tsai, L.~Visinelli, M.~Afrin, A.~Allahyari, P.~Bambhaniya, D.~Dey, S.~G.~Ghosh and P.~S.~Joshi, \textit{et al.}
Class. Quant. Grav. \textbf{40}, no.16, 165007 (2023)
doi:10.1088/1361-6382/acd97b
[arXiv:2205.07787 [gr-qc]].

\bibitem{Jolliffe}
Jolliffe, Ian T and Cadima, Jorge, ``Principal component analysis: a review and recent developments,'' Phil.\ Trans.\ R.\ Soc.\ A.\ {\bf 374}, 20150202 (2016).

\end{thebibliography}
\end{document}